\documentclass[usenatbib]{mn2e}  
\usepackage{graphicx}
\usepackage{txfonts}
\input{epsf}
\newcommand{\apjl}{ApJL}
\newcommand{\apj}{ApJ}

\newcommand{\aj}{AJ}
\newcommand{\mnras}{MNRAS}

\newcommand{\apjs}{ApJS}

\newcommand{\aap}{A\&A}
\newcommand{\aaps}{A\&ASS}

\begin{document}
   \title[SDSS DR7 colour gradients]{Colour gradients within SDSS DR7 galaxies: 
          hints of recent evolution}

   \author[Gonzalez-Perez, Castander \& Kauffmann]{
     \parbox[t]{\textwidth}{
     \vspace{-1.0cm}
     V. Gonzalez-Perez$^{1,2}$, F. J. Castander$^2$,
     G. Kauffmann$^3$}\\
  $^1$Institute for Computational Cosmology, Department of Physics, 
University of Durham, South Road, Durham, DH1 3LE, UK\\
  $^2$Institut de Ci\`{e}ncies de l'Espai (CSIC/IEEC), F. de Ciencies, Torre C5 Par 2a, UAB, Bellaterra, 08193 Barcelona, Spain\\
  $^3$Max-Planck Institut f$\ddot{u}$r Astrophysik, D-85741 Garching, Germany
          }

\maketitle

\begin{abstract}
  The evolutionary path followed by a galaxy shapes its internal
  structure, and, in particular, its internal colour variation. 
  We present a study of the internal colour variation within galaxies
     from the Seventh Data Release of the Sloan Digital Sky Survey
     (SDSS DR7). We statistically study the connection between the
     internal colour variation and global galactic properties, looking
     for hints of the recent galactic evolution.
  Considering only galaxies with good photometry and spectral
  measurements, we define four luminosity-threshold samples within the redshift range $0.01<z<0.17$,
  each containing more than $48000$ galaxies. Colour gradients are calculated for these galaxies from the surface brightness
  measurements provided by the SDSS DR7. Possible systematic effects in their determination have been analysed.
   We find that, on average, galaxies have redder cores than
   their external parts. We also find
   that it is more likely to find steep colour gradients among late-type
   galaxies. This result holds for a range of classifications based on both morphological and spectral characteristics. In fact, our results relate, on average, steep colour gradients to a higher presence of young stars within a galaxy. Our results also suggest that nuclear activity is a marginal driver for creating steep colour gradients in massive galaxies. We have selected pairs of
   interacting galaxies, with a separation of $5$ arcsec, in projected radius,
   and a difference in redshift of $100\,km/s$,
   finding that they present steeper gradients than the average
   population, skewed towards bluer cores. Our analysis implies that colour gradients can be
     useful for selecting galaxies that have suffered a
   recent (minor) burst of star formation.
\end{abstract}
   
\begin{keywords}
 galaxies: fundamental parameters, statistics, structure
\end{keywords}


\section{Introduction}

The subtleties of galaxy formation remain to be understood. The
study of the internal structure of galaxies can help us to understand
how they formed and evolved. Here we study the internal colour
variation across galaxies, quantified through their colour
gradient. Interest in internal colour variation has grown
enormously in the last five years, especially due to the availability
of both shallow large surveys such as the Sloan Digital Sky Survey
\citep[SDSS, ][]{york:00}, that can provide statistically significant
samples of galaxies, and deep surveys, that provide high
resolution images of galaxies at $z>0.5$, such as those performed with
the Hubble Space Telescope \citep[HST; e.g.][]{menanteau:05, ferreras09}. Previous work has used the colour gradient as a measure of the
galactic internal colour variation, which has proved to be very useful for
exploring galactic properties, due to its
tight dependency on the galactic distribution of metallicity, stellar
age and dust \citep[see e.g.][]{wu:05,michard:05,tim10,tortora10}. This,
together with the accessibility of gradient measurements, since only
photometry is needed, makes the colour gradient a useful tool to obtain statistical results for wide samples of galaxies, that will be interesting to relate to different theoretical aspects from models of galaxy formation. 

In general, galaxies tend to have cores that are redder than their external
parts \citep[e.g.][]{peletier:90}. In this work we study the variation
within nearby galaxies of the following optical colours: $(g-i)$, $(g-r)$, $(r-i)$,
$(i-z)$ and $(r-z)$. Both $(g-r)$ and $(g-i)$ rest frame colours are
most sensitive to the intensity of the spectral break
that appears around $4000$ {\AA}, due to the accumulation of resonant absorption
lines from metals in different stages of ionisation
\citep[e.g.][]{hamilton85}. This break correlates with the temperature
of the dominant stellar population, increasing for cooler stars, but
it does not depend strongly on the galactic metallicity. Unlike
late-type galaxies, early-type ones are dominated by
K-type stars, showing a rather steep change in flux from g to r or i-band. If we consider that these colours are mainly related
to the age of the galaxy then the redder core will imply that it is also older.

The other three colours, $(r-i)$, $(r-z)$ and $(i-z)$, in the rest
frame directly depend on the proportion of main sequence to red
giant branch or older stars \citep{bruzual93}. Thus, we could simply record these
colours as depending on the relative amount of old and young stars in
the galaxies, though this is less clear for $(i-z)$ and has a smaller
dynamic range in the case of the $(r-i)$ colour. Thus, we observe the same tendency as before: redder cores imply that in general galaxies tend to have older stars in their centres. 

However, these implications should be taken with caution for two
  main reasons: i) Global colours can be affected by marginal young
  populations \citep{li07}. ii) Colours
within different galactic annuli do not necessarily agree well with global
ones. Besides, broad band colours show a degeneracy in their dependence with stellar population age and metallicity \citep[e.g.][]{worthey94,charlot96}. Indeed, several studies \citep[see
  e.g.][]{tamura04,wu:05,lb09,tortora10} have shown that
metallicity and not age gradients are the principal origin of the colour
variation. In the case of early-type galaxies, age gradients have been
measured to be almost flat or positive while metallicity gradients are
negative, i.e. early-type galaxies tend to have younger, higher
metallicity cores \citep[see
  e.g.][]{sanchez-blazquez,clemens09,lb09,tim10}. Late-types galaxies are more complex, showing a broad variety of age and also metallicity gradients, due to the interplay between dust content, stellar feedback, bursts of formation and stellar migration due to the instabilities that trigger the formation of bars but also the spiral arms. \citet{macarthur09} observed that spiral galaxies have bulges with positive age gradients, while their disks show negative ones, but there are examples of spiral galaxies with negative age gradients for most of their extension, except for the most external radii \citep{vlajic09}. 

Metallicity gradients within galaxies can be produced through
different mechanisms. Winds generated by supernovae \citep{larsonSN:74}
or active galactic nuclei (AGN) \citep{begelman91} can provide the metallicity enrichment
of the inter stellar or even the inter galactic medium. Due to the multiple
phases that compose this medium the enrichment does not necessary
happen in a symmetrical way, which can produce strong metallicity
gradients and thus colour gradients. Galaxies with a significant age
gradient can present very shallow or even inverted colour gradients
\citep{hinkley:01}, since age gradients usually arise from localised
regions of star formation, generally resulting in galaxies with bluer cores than
their external parts. The presence of a diffuse component of dust
could also explain some colour gradients. However, this
has a marginal impact, specially for early-type galaxies, which are almost dust free
\citep{michard:05, wu:05}. In summary, colour gradients in galaxies are
likely to be related to the different star formation histories
followed, which depend on the way gas is distributed and collapses after accretion and mergers.  

Many studies of colour gradients try to clarify the dependence on the
star formation history of galaxies. This aspect is of particular
interest since it could help us to understand how galaxies form and
evolve. The recent study by \citet{suh10} of early-type galaxies drawn
from the SDSS DR6 reports a tight correlation between the existence of
steep colour gradients and ongoing residual star formation. This study
of galaxies at $z<0.06$ suggests a relation between elliptical
galaxies with bluer cores presenting global bluer colours than
average. \citet{ferreras09} found the same relation for spheroidal
galaxies observed by the HST at $0.4 < z < 1.5$. \citet{lee08} found that steeper colour gradients appear within star
forming galaxies, in both late and early-types. A similar result was
found in the recent study by \citet{tortora10}, where older galaxies
are observed to have systematically shallower age and metallicity
gradients, and thus, colour gradients. The colour
gradients of spiral galaxies appear to be dominated by the fact that, on average, their bulges are redder than their disks. However, this is an oversimplification for many late-type galaxies \citep[see e.g.][]{bakos08}.

Cosmological simulations do not yet have sufficient resolution to make
predictions for colour gradients. However, semi-analytical models
which include a bulge-disk decomposition and spatially resolved star
formation, could provide some useful insight for this particular
problem \citep{stringer07}. Nevertheless, in general, gas-dynamical
simulations with high enough resolution to follow physical processes
within galaxies are needed in order to produce theoretical predictions
for the internal characteristics of galaxies. ``Monolithic
  collapse'' models can produce elliptical galaxies
with colour gradients that agree with
observations if a certain scatter is allowed for the star formation
efficiency \citep{pipino10}. These monolithic models also predict steeper
gradients for more massive galaxies, something that is observed up to $M_* \sim 3.5\times 10^{10}\,
M_{\odot}$ \citep{spolaor09,tim10,tortora10}. For more massive galaxies, the contribution of mergers provides a larger scatter that smears out such correlation. \citet{kobayashi:04} used chemo-dynamical simulations of merging galaxies to understand elliptical galaxy formation. The resultant galaxies had metallicity gradients consistent with observations. The predicted gradients depend strongly on the relative properties of the progenitor galaxies, i.e., on the particular formation process of a galaxy. 

In this paper, we study the largest data set to date, which allows
us to make a statistically robust analysis of the internal colour
variation within nearby galaxies. We study four luminosity-threshold
samples, defined in section \S\ref{sec:dr7sample}. We obtain colour
gradients directly from the SDSS photometric data and we (K+e) correct
global magnitudes using spectral energy distributions (SEDs)
generated with {\sc pegase} \citep{fioc:97}. This process is described in section \S\ref{sec:k+e_cor}. In section \S\ref{sec:results} we analyse the colour gradients distribution and we also study the possible correlations between the colour gradient and global galactic properties, including galactic types and redshift. Finally we summarise our results and conclusions in section \S\ref{sec:conclusions}.

In this study we use Petrosian magnitudes for setting the sample limits. Apparent magnitudes are corrected from the
extinction due to our galaxy, but not to their internal
reddening. Model magnitudes are used only for global galactic
colours. All
magnitudes used in this paper are in the AB system, unless otherwise
specified. We assume a flat cosmology, $\Omega_{K}=0$, with a critical
matter density of $\Omega_m=0.3$ and a Hubble constant, $H_0=100\, h\,
km\, s^{-1}\, Mpc^{-1}$ with $h=0.7$ \citep{wmap1}.


\section{The data}\label{sec:dr7sample}

Our sample is drawn from the SDSS
DR7 \citep{dr7} spectroscopic main galaxy targets. The SDSS DR7 has covered spectroscopically an
area of $9380\, deg^2$, providing a redshift accuracy of $30\, km/s$
rms for the main galaxy sample. The SDSS main spectroscopic sample is
described by \citet{strauss:02}. Galaxies within this sample have
$r<17.77$ (Petrosian
magnitude after correcting for
galactic extinction) and a median redshift of $z\sim 0.1$.

From the SDSS main spectroscopic sample, we further select those galaxies that, according to the SDSS pipeline flags, have a reliable determination of
their redshift, a good photometry\footnote{By good photometry we mean
  here that according to the SDSS pipeline flags the galaxies are
  detected at more than $5\sigma$ level, with all their pixels checked
  for possible flux peaks, not being a blended object, without too many interpolated pixels, without pixels hit by a cosmic ray and without a faulty sky subtraction.}
in all g, r, i and z bands, have a good determination of their Petrosian radius and a complete surface brightness
profile measurement. In order to
avoid those galaxies most affected by aperture effects, a lower limit
to the galaxy redshift has been imposed, $z>0.01$. 

After applying the above restrictions we found that one of the
galaxies appeared multiple times due to different deblending of
objects at successive scans. After taking out this galaxy our initial sample contains $339711$ galaxies.

As it will be detailed in \S\ref{sec:howgrad} we quantify the
  internal colour variation using the SDSS DR7 measurements of surface
  brightness averaged within circular concentric annuli. At most there
  are $15$ measurements, though, in most cases there are less due to
  the low surface brightness of many galaxies. The outer radii of these
  annuli are fixed and range from $0.22$ to $258.38$ arcsec, growing approximately
  exponentially\footnote{For further details we refer the reader to
    the SDSS help page
    http://cas.sdss.org/dr7/en/help/browser/constants.asp?n=ProfileDefs
    and the document on profiles written by R. Lupton \& J. Gunn: 
    http://www.astro.princeton.edu/$\sim$rhl/photomisc/profiles.ps }.

We make use of the stellar masses, star formation rates, some
spectral indices and classification provided in the SDSS
MPA/JHU\footnote{http://www.mpa-garching.mpg.de/SDSS/ .}
collaboration database \citep{kauffmann_mass:03, brinchmann:04}, which
has been updated for the SDSS DR7. These parameters are obtained by
extrapolating to the full galaxy the spectroscopic information
collected within the $3$ arcsec diameter fibre set around the galactic
centres. The extrapolation is done taking into account the global galactic colour and the effects due to
dust. This extrapolation does not consider the colour gradients within galaxies, which could include a systematic effect on the determination of global properties.

\subsection{(K+e)-correction}\label{sec:k+e_cor}

\begin{figure}
\includegraphics[width=9.cm]{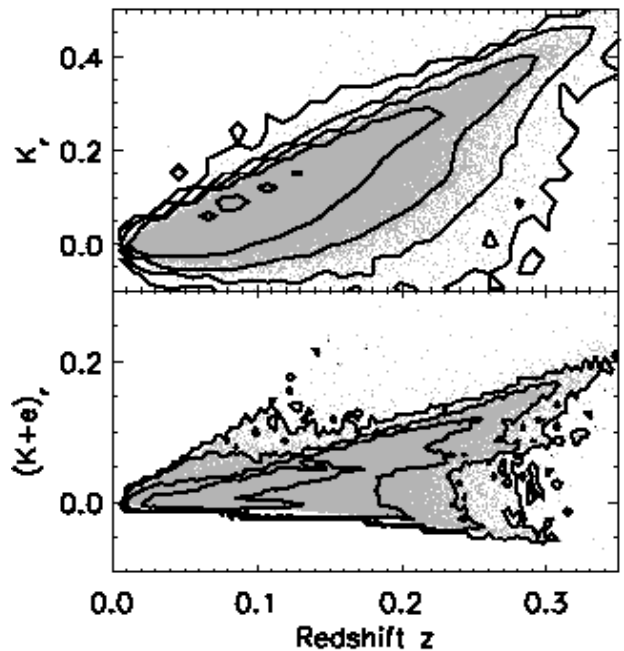}
\caption{ K correction (top) and (K+e) correction (bottom) in
    the r-band as a function of redshift estimated by fitting {\sc
      pegase} stellar population models to SDSS DR7 galaxy photometry. Contour levels represent an
    increase of a factor of ten and they are plot over single galaxy
    data points.
}
\label{fig:kcor}
\end{figure}

The (K+e)-correction accounts for two effects. One, the K-correction,
is the shift of the filter in the galaxy's rest frame with redshift
due to observing with a fixed band \citep{oke68}. The other, the e-correction,
accounts for the change in luminosity due to ageing of existing stellar
populations or new star formation that galaxies experience up to a
reference redshift, generally $z=0$ as in our case, from the epoch at which they are observed \citep{poggianti97}. Both
effects vary between the different bands. Within the SDSS bands we
obtain larger corrections for bluer bands \citep{fukugita:95}. In order to make sure that we are comparing similar spectral regions from different galaxies
we need to apply the (K+e)-correction.

The (K+e)-correction is evaluated fitting the galaxy photometry to the
best model within a grid of SEDs generated with the population
synthesis models {\sc pegase} \citep{fioc:97}. The grid is made using the default template SEDs produced within {\sc pegase}, matching the colours observed for different Hubble galactic types at
$z\sim0$, and includes a galaxy formed in a $1\, Gyr$ long burst, an elliptical and a lenticular galaxies, spiral galaxies: Sa, Sb, Sbc, Sc, Sd and three irregular galaxies with different formation redshifts. 

There are six galaxies in our initial sample with colours that are not
well fitted by the synthetic ones. These are left out of the
sample. Since these objects represent $0.002$\% of the sample we do
not apply any correction for their omission.

Both K and (K+e) corrections are shown in Fig. \ref{fig:kcor} for all
galaxies in our initial sample. These corrections are applied to the global
colours of galaxies in our sample. The K-correction alone tends to be
larger for higher redshifts, but the dispersion increases at the same time.  This behaviour has been found in other works \citep[see e.g.][]{blanton:03,fukugita:95}. 

We can observe in Fig. \ref{fig:kcor} that the (K+e)-correction shows
a weaker dependency on redshift than the K-correction. This was also
found in previous studies, such as \citet{poggianti97}. The reason for
this flattening is that as galaxies evolve, their luminosity decreases
until a new burst of star formation occurs, being superimposed to an already old stellar population. 

We checked the possibility of calculating the (K+e)-correction
individually for different galactic radii. For this purpose, we obtained magnitudes in circular concentric annuli for
each galaxy, from the averaged surface brightness measured by the SDSS
pipelines. 
However, colours in individual annuli are noisier than global ones and the fitting procedure basically fits noisy features that result in obtaining unrealistic SEDs for each annulus.
Therefore, we obtain the global (K+e)-correction for each galaxy, which
presumably is an average of the correction to different annuli. The use of an average (K+e)-correction could introduce spurious colour
gradients. These are difficult to evaluate, since they depend on the intrinsic metallicity and
age gradients of each galaxy, 
higher signal-to-noise data would be needed to properly compute
these corrections as a function of radius and evaluate the effect of observing different radii at a given projected annulus.
Applying a global corrections implies that the resultant
colour gradients would not be affected, since we are only moving the
zero point on the colour vs. radius relation. This implies that
similar colour gradients will be found either applying the
(K+e)-correction or only the K-correction. However, applying one or
the other changes very slightly the defined luminosity-threshold
samples, and has a definite impact on the global colours associated to
each galaxy.

We will restrict our study to galaxies seen up to
such a redshift that the (K+e)-corrections would not be large. Rather
than the actual value of the (K+e)-correction, the problem resides in
the higher spread among template SEDs found at higher redshifts, which
can introduce larger errors in the estimated
correction. For the majority of
galaxies up to $z<0.17$, the (K+e)-corrections is well below a value
of $0.6$ in the g, r, i and z-bands. For the same range, the
K-correction stays below $0.8$ in all bands. There is another reason
for not considering galaxies at the highest redshifts covered by the
survey. While the distribution of the K-correction is smooth, it is
clear that our estimate of the (K+e)-correction start to be almost
discretised for galaxies with $z\ge 0.25$. This effect is not
  present at $z<0.17$. In order to correct
  galaxies at $z>0.17$ we need to enhance our grid of
  model SEDs, including different formation redshifts for not only
  Irregular galaxies but for all the
  Hubble types included as templates. This change should be done
  carefully, in order to reproduce the observed fluxes of galaxies at
  different evolutionary stages.

Taking into account the two mentioned sources of uncertainty we will restrict our study to galaxies with $z<0.17$.

\subsection{The luminosity-threshold samples}

\begin{center} 
\begin{table}
\begin{tabular}{c c c c}
\hline
Name & Absolute magnitude   & Redshift  & No. galaxies
\\
\hline
{\it S19} & $M_{r}-5logh<-19.$ & $0.01<z<0.072$ & 48728
\\
{\it S20}  & $M_{r}-5logh<-20.$ & $0.01<z<0.111$ & 82838
\\
{\it S20.5} & $M_{r}-5logh<-20.5$ & $0.01<z<0.1375$ & 86818
\\
{\it S21} & $M_{r}-5logh<-21.$ & $0.01<z<0.169$ &  66177
\\
\hline
\end{tabular}
\caption{The given name, absolute magnitude range, redshift range and
  final number of galaxies to be studied within the defined luminosity-threshold samples.}
\label{tab:sample}
\end{table}
\begin{figure}
\includegraphics[width=9.cm]{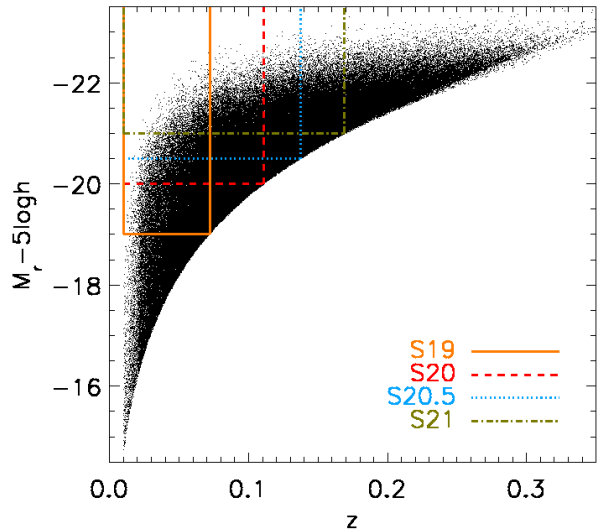}
\caption{ Redshift versus absolute magnitude in r band for the
    selected SDSS DR7 galaxies. The different luminosity-threshold samples are
    enclosed by solid lines for the {\it S19} sample, dashed for the {\it
    S20}, dotted for the {\it S20.5} and dash-dotted for the {\it S21}.}
\label{fig:sample}
\end{figure}
\end{center}

We have divided our initial sample of galaxies with $z<0.17$ into four
luminosity-threshold samples. These samples, though well defined, are
not volume limited since they miss a cut in redshift
corresponding to the bright limit of the survey. Fig. \ref{fig:sample} shows the boundaries of each
sample superimposed on the distribution of the absolute magnitude
versus redshift for all galaxies. Absolute magnitudes have been obtained from the
(K+e)-corrected Petrosian magnitudes in the r-band. Table \ref{tab:sample} summarises the characteristics of the luminosity-threshold samples. We will refer to
each sample by their absolute magnitude cut, as indicated in the first
column of Table \ref{tab:sample}.

We find that some of the galaxies within the {\it S21} sample have a
half light radius $R50<0.7$ arcsec, i.e., inside the first two annuli with surface brightness measurements. As explained in \S\ref{sec:howgrad} these two annuli are not considered
in the calculation of the colour gradients. Thus, for those galaxies within the {\it S21} sample we will be calculating colour gradients with only the outer
half of the galactic light. For this reason, we will concentrate on
the other three samples at $z<0.14$, and results from the {\it S21} sample will be considered with caution.

 As explained in the next section
\S\ref{sec:howgrad}, to obtain colour gradients, galaxies should
have their surface brightness measured in at least four annuli,
without considering the two innermost ones. Among our samples we find
two galaxies which have their surface brightness measured in too few
annuli, one within the sample  {\it S20.5} and the other within the
{\it S19} one. These
two galaxies lack a colour gradient measurement and are not considered further.  

As mentioned previously, we make use of some of the quantities derived
from the SDSS database by the MPA/JHU collaboration. The cross
identification using the position coordinates of the galaxies is not
perfect. Fewer than $5$\%  of the galaxies in our initial samples do not appear to have properties measured
by the MPA/JHU collaboration. Only when using the MPA/JHU derived parameters we will be neglecting those galaxies.

\subsection{Quantifying the internal colour variation through colour gradients}\label{sec:howgrad}

Different ways to compute the internal colour variation can be
  found in the literature. Some examples are: the slope
  of colour vs. normalised radius or its logarithm \citep[e.g.][]{hinkley:01,michard:05,tortora10}, the difference between
  colours outside and inside certain radius \citep[e.g.][]{menanteau:05,park05}, the
  comparison between model and fibre colours when working with
  galaxies from the SDSS \citep[e.g.][]{roche09} and the colour variation
  averaged to all the pixels in an image \citep[e.g.][]{menanteau:01}.

In the early stages of this work we used the SDSS DR2 to estimate the best way to quantify the internal colour variation. We studied the internal colour
variation within a sample of galaxies with a concentration index $C\ge 2.63$ and $z<0.1$. The concentration index is defined as the ratio between two radii enclosing different amounts of light flux. The SDSS definition uses the radii containing $90\%$, {\it R90}, and $50\%$, {\it R50}, of the total galactic light. \cite{strateva:01} and \cite{shimasaku:01} found that
$C=2.6$ is the boundary that separates early from late-type galaxies, according to both their morphological and spectral
features. The actual value of this boundary can vary depending on level
of restriction needed for contaminants within one or the other type of
galaxies

After thoroughly testing various methods using the quantities provided by the SDSS database, we reached the conclusion that the surface brightness profile appears to be the best way to quantify the internal
colour variation in galaxies as opposed to magnitude based estimations,
e.g. comparing the {\it fibre} or {\it PSF} colours with the global
galactic colours. 

The SDSS provides measurements of surface brightness averaged
within circular concentric annuli at different galactic radii. We
obtain colour gradients from these measurements. The two innermost
annuli, $R<0.7$ arcsec, are excluded from the calculation, since they are
the most affected by seeing (the median PSF width in the r band for
the SDSS DR7 is $1.4$ arcsec). Previous studies, such as that from \citet{tamura03} have reported that the internal colour variation is diminished if the annuli most affected by seeing are considered for obtaining the surface brightness profile.

From the SDSS averaged surface brightness profile we get the
cumulative one. This last one is fitted to a taut spline\footnote{A
  taut spline is a piecewise cubic interpolant for which the first and
  last internal knots are not used. See \citet{deboor:78} for a full
  description.}, using a fourth order polynomial. Thus we need galaxies
to have measurements of average surface brightness in at least four
annuli. The interpolated surface brightness points are
differentiated. We then convert the resulting surface
brightness into
magnitudes. We do this last step only for the radii corresponding  to
the outer ones from the initial annuli. These magnitudes are corrected with the global
(K+e)-correction obtained for the whole galaxy. Colour gradients are
finally obtained by calculating the slope of the straight line that
best fits the colours, $(m_{1}-m_{2})$, at different galactic radii (those
for which a surface brightness profile is provided in the SDSS) versus the radii of the annuli normalised by the galaxy half light radius in the r-band,
$R50$:
\begin{equation}\label{eq:grad}
\bigtriangledown _{m_{1}-m_{2}}=\frac{\bigtriangleup(m_{1}-m_{2})}{\bigtriangleup(R/R50)}
\end{equation}

The normalisation in radius is done in order to allow us to compare
galaxies with different light distributions and sizes. The colour
  gradients of all the
  galaxies within the four studied luminosity-threshold samples have
  been obtained from the surface brightness profile out to the annulus
  containing the radius $R=2\times R50$. Thus, the radial coverage is consistent with previous
studies. A further analysis on the effect of radial coverage is made
in \ref{a_rad}. In order to allow a fair comparison with recent
literature we have also obtained colour gradients from the variation
of internal colour with the logarithm of the normalised radius (colour
per radial decade).

When computing the errors on the measured gradients we use a Monte Carlo method to take into account the
observational errors in both the surface brightness and $R50$
measurements. Surface brightness values
within a galaxy are related, with one value influencing the values at
outer radii. In order to simplify the calculation we use a compromise
 by effectively considering the
surface brightness values as independent and restricting the random values, obtained with the Monte Carlo
method, within
$1\sigma$. Thus, in order to include the observational errors, we repeat the whole process, except for the calculation of
the (K+e)-correction, $1000$ times for each galaxy using
each time a random value extracted from the error distribution for the
variables within $1\sigma$ of the measurements, assuming that the
initial errors have a Gaussian distribution. 

We obtain the probability distribution from the Monte Carlo runs
making use of a {\it simple estimator}\footnote{The simple estimator,
  f, is basically a weighted histogram without a dependency on the
  selection of the 
  first bin, though it maintains a dependency on the step size, h:
\begin{equation}
f(x)=\frac{1}{n}\sum_{i=1}^{n}\frac{1}{h}w\bigg(\frac{x-x_{i}}{h}\bigg)
\end{equation}
Here the weight, w, is $0.5$ if $|x-x_{i}|<1$ and $0$ otherwise.} and then we calculate the expected value of the colour gradient and its variance.

\section{Results}\label{sec:results}

Here we explore the origin of the colour gradient distribution in
connection with the different classifications of galaxies into either
early and late types or star forming, passively evolving and hosting nuclear activity.

\begin{figure}
{\epsfxsize=8.5truecm \epsfbox[68 17 546 473]{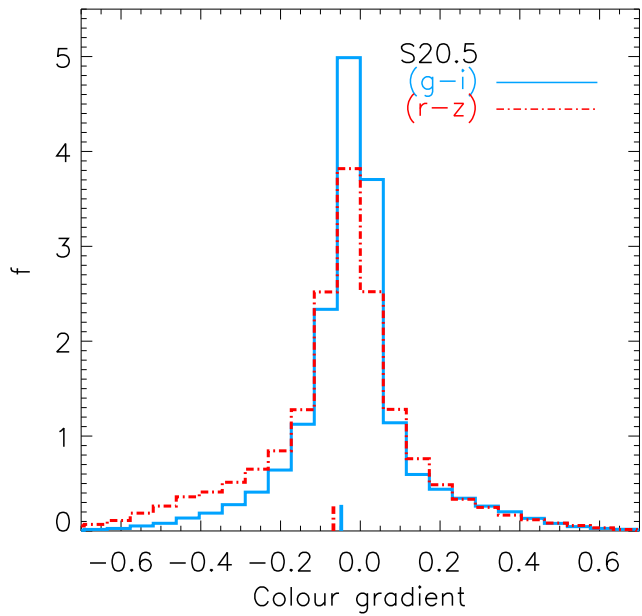}}
\caption{ Distribution of $(g-i)$ (solid line) and $(r-z)$ (dot-dashed line) colour gradients for galaxies within the sample
    {\it S20.5}.  Median colour gradient values are shown as short lines of the same type as their corresponding distribution. Histograms are normalised to give unit area underneath.}
\label{fig:allgrad}
\end{figure}

We have calculated the gradients for the $(g-i)$, $(g-r)$, $(r-i)$,
$(i-z)$ and $(r-z)$ colours. Fig. \ref{fig:allgrad} shows the
distribution of the $(g-i)$ and $(r-z)$ colour gradients for galaxies in
sample {\it S20.5}. We have found that the distributions of the colour gradients
in the $(g-r)$ and $(g-i)$ colours are similar. The distributions of the colour gradients in the $(r-i)$, $(r-z)$ 
and $(i-z)$ colours are also similar between them. As described in the
introduction this is due to the similarity in the spectral characteristics
that these colours probe. 

\begin{figure}
{\epsfxsize=8.5truecm \epsfbox[55 7 622 309]{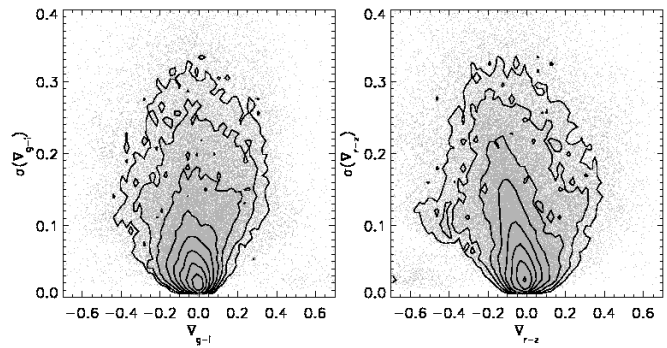}}
\caption{ Distribution of colour gradients errors for $(g-i)$,
    (left) and $(r-z)$ (right) for galaxies in the {\it S20.5} sample. Each contour represents a factor of two change in density.}
\label{fig:err_grad}
\end{figure}

Fig. \ref{fig:err_grad} presents the dependency of the variance in the
colour gradient with the actual values of the $(g-i)$ and
$(r-z)$ colour gradients. From this figure we learn that galaxies with
redder cores (negative colour gradient) measured from the
$(r-z)$ colour gradients, tend to have larger errors, due to the
z band being the noisiest among the four in this study. The same is
not true for the $(g-i)$ colour gradients. It can be seen in
Fig. \ref{fig:err_grad} that steeper gradients tend to have larger
errors, for both colour gradients.

Fig. \ref{fig:allgrad} shows that both the $(g-i)$ and $(r-z)$
colour gradients for sample {\it S20.5} have median values that are
similar: $-0.048$, $-0.067$, respectively ($-0.187$, $-0.277$, when
defining colour gradients as colour per decade in normalised
radius, in decimal logarithmic units). However, the distribution for
the  $(r-z)$ colour gradient presents a clear longer negative tail. This slight difference is
likely to be due to the fact that redder stars contribute more in the $(r-z)$ than in the $(g-i)$ colour, giving rise to
steeper gradients. However, the tilted distribution of the
$(r-z)$ colour gradient variance shown in Fig. \ref{fig:err_grad},
would also be responsible, at least in part, for this difference.

We have also compared the distribution of colour gradients for the
different luminosity-threshold selections defined in Table
\ref{tab:sample}. There is not a clear change in the shape of the distributions
for the different samples, just different numbers of
galaxies. The change in median values from one sample to another
  is rather small: $-0.050$, $-0.049$, $-0.044$ for the median $(g-i)$
  colour gradient of samples {\it S19}, {\it S20} and {\it S21},
  respectively.  When defining colour gradients as colour per decade in normalised
radius we obtain the following median values $-0.185$, $-0.194$,
$-0.167$ for samples {\it S19}, {\it S20} and {\it S21}, respectively.
These last values are roughly in agreement with the
\citet{tortora10} study, where they found a non monotonic dependency
of colour gradients with magnitude, when galaxies fainter than $M_r=-20$
were considered.

In appendix \ref{ap-sys} we analyse some of the systematic effects that could affect
the distribution of colour gradients, including the ellipticity,
the amount of light considered in the calculation and the effect of
having close pairs of galaxies. Considering the result found in
appendix \ref{ap-sys} and the similar distributions for all the luminosity-threshold samples considered in this
work, we can conclude that the results for the colour gradient
distribution are robust.

Returning to Fig. \ref{fig:allgrad}, we can clearly see there that galaxies
tend to have small internal colour variation, resulting in quite flat
colour gradients. Nevertheless, the median of the distribution is
negative, implying that, in general, galaxy centres are redder than their
outer parts. This agrees with previous
studies using smaller samples of
galaxies \citep[see e.g.][]{peletier:90,michard:05,lee08,
  tortora10}. As described in the introduction, Fig. \ref{fig:allgrad}
shows then that galaxies tend to have more metal rich cores that might
also be hosting older stellar populations. Though, age gradients
present larger dispersion than metallicity ones \citep{macarthur09,vlajic09,tim10,tortora10}.    

A remarkable feature of Fig. \ref{fig:allgrad} is that the colour
gradient distributions presents a
single peak. This result could be indicating that galaxies with
different structural characteristics, late and early type galaxies,
have undergone similar star formation histories. However, the
distribution of colour gradients presented in Fig. \ref{fig:allgrad}
is much broader than a Gaussian. We have checked whether this is due
to the observational errors using Monte Carlo simulations. We have
found that taking into account the colour gradients variance, we still
obtain a distribution narrower than the observed one. Thus, we
demonstrate that the width and shape of the colour gradient
distribution cannot be explained by the error distribution alone and
that galaxies do indeed present different intrinsic colour
gradients. We find the colour gradient distribution to be best
described as the superposition of the distributions of two populations
of galaxies.

We will explore further this last aspect in \S\ref{sec:types} and
\S\ref{sec:agn}. Before, we analyse the potentially different
nature of those galaxies with extreme positive and negative colour gradients.

\subsection{Positive and negative colour gradients}\label{sec:extreme}

Here we study whether extremely steep colour gradients occur only in
certain types of galaxies.

\begin{figure}
{\epsfxsize=8.5truecm \epsfbox[12 14 557 495]{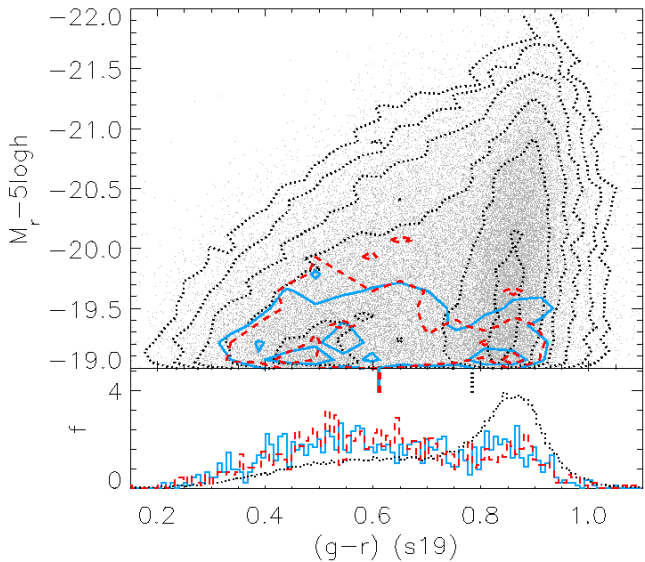}}
\caption{ The upper panel shows the absolute magnitude in the r-band versus the global colour
  $(g-r)$ for galaxies within the sample {\it S19}. Contours
  increase by factors of two in density, illustrating the behaviour of
  galaxies within different ranges of the $(g-i)$ colour gradient
  distribution. Dotted contours correspond to galaxies with $(g-i)$
  colour gradients within the $2\, \sigma$ limits of the total
  distribution. Solid/Dashed contours present the distribution of
  those galaxies with their $(g-i)$ colour gradients above/below the
  $2\, \sigma$ limit of the distribution, i.e., with bluer/redder
  cores than the average trend. The lower panel shows, in the
    corresponding line types, the medians and the distributions of the
    global $(g-r)$ colour for the three subsamples selected according
    to the $(g-i)$
  colour gradients values. Histograms are normalised to give unit area underneath.}
\label{fig:color_gr}
\end{figure}

\begin{figure}
{\epsfxsize=8.5truecm \epsfbox[12 14 557 495]{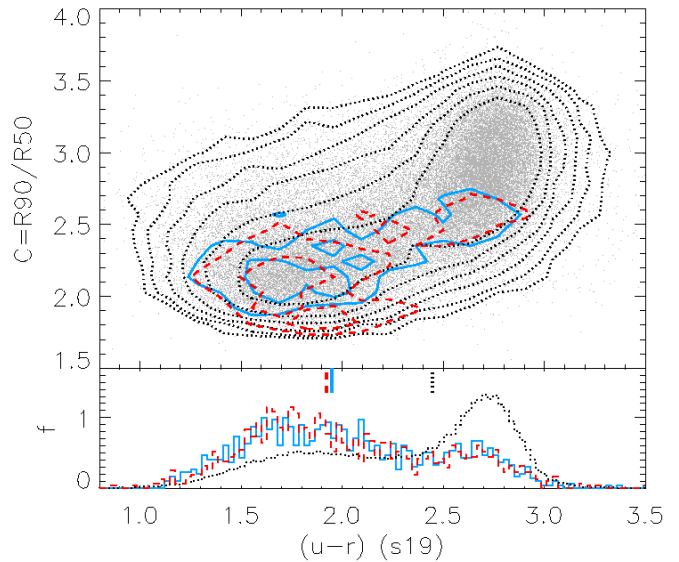}}
\caption{ Similar to Fig. \ref{fig:color_gr} but for $(u-r)$ vs concentration index. }
\label{fig:color_ur}
\end{figure}

In Fig. \ref{fig:color_gr} and  Fig. \ref{fig:color_ur} we explore the
possibility for galaxies with positive and negative colour gradients
to separate in a colour-magnitude or colour-concentration parameter
space. In both plots, density contours correspond to the behaviour of
three subsamples separated by their $(g-i)$ colour gradient values. We
have set as boundaries the $2\, \sigma$ values corresponding to the
$(g-i)$ colour gradient distributions of each sample. In this way we
separate those galaxies with extreme colour gradients from the global
trend. 

Fig. \ref{fig:color_gr} shows the absolute magnitude versus the global $(g-r)$
colour, while Fig. \ref{fig:color_ur} presents $(u-r)$ versus light
concentration. Both parameter spaces were explored by
\citet{blanton:03} and \citet{driver06}, respectively. These studies
showed the bimodality of galaxies, which separate in these spaces into
early and late-type.  In
the \citet{blanton:03} observations, the bimodality of SDSS galaxies is
clear for galaxies fainter than $M_{0.1\, r}\sim -20$ (absolute
magnitude at $z=0.1$). A similar trend is found here for the faintest
sample {\it S19}. The brighter samples, {\it S20, S20.5, S21}, do not show a clear bimodality due to the
magnitude cut. 

In Fig. \ref{fig:color_gr} those galaxies with very steep colour
gradients only appear at the faint end, corresponding to the region
with the largest dispersion, but also with the largest errors in
magnitude. Galaxies with very steep positive and negative colour are
distributed similarly in Fig. \ref{fig:color_gr}. 

Fig. \ref{fig:color_ur} shows that galaxies with extreme colours
neither separate in this parameter space.  The same is seen when
  defining the colour gradient as colour per decade in normalised
  radius. Nevertheless, Fig. \ref{fig:color_ur} shows a remarkable
behaviour: the galaxies with the steepest colour gradients are
  not galaxies with high concentration indices. Galaxies with high
concentration indices are generally passively evolving with minimal
content of gas. Since few galaxies with high concentration index
  show steep colour gradients, then one expects that these steep
  colour gradients are more likely related to star formation. We will
further explore this in \S\ref{sec:types}.

From both Fig. \ref{fig:color_gr} and  Fig. \ref{fig:color_ur} it is clear that
galaxies with steep colour gradients, independently of their sign, appear in larger numbers towards
the blue end of both the $(u-r)$ and $(g-r)$ global colour
distributions. Both figures show that redder and bluer cores appear
within galaxies with similar global colours. However, for
brighter samples, we find clearly that redder cores appear, on
average, within redder galaxies than those with bluer cores. In
particular, for the sample {\it S20.5} those galaxies with extremely
blue (red) cores have median global colours: $(g-r)=0.82\, (0.85)$, $(u-r)=2.3\, (2.4)$, while those
galaxies with flatter colour gradients present the following median
global colours: $(g-r)=0.94$, $(u-r)=2.8$. 

We have also explored whether galaxies with very steep gradients
appear separated when looking to parameter spaces that depend on age
or metallicity against concentration index. In particular we have
studied the variation with light concentration of the $D4000$ spectral
index and the Lick index $H_{\beta}$, both of which are mostly sensitive to
age. We have also compared concentration to the Lick indices $Mg2$ and
$Fe5335$, both closely related to the global metallicity of the
galaxy.

We observe that both $Fe5335$ and $H_{\beta}$ present values that do not appreciably change with light concentration, which prevents the
separation of galaxies in such parameter space. We find a similar bimodality in the parameter space $D4000$ vs. $C$, to that reported by
\citet{kauffmann_tot:03} for the SDSS DR2. Moreover, we find that galaxies also separate into early and late-types in the parameter
space $Mg2$ vs. $C$. We find again that galaxies with very steep
gradients appear to have a wide range of indices values, basically
covering the same range as the global trend. It is also interesting to
note that this is independent of the steep colour gradients being
positive or negative. The lack of a clear locus occupied by those
galaxies with steep gradients in the different studied parameter spaces suggest that the existence of extreme colour
gradients depend weakly on the global age or metallicity of
galaxies. 

The studied indices are measured within the SDSS DR7 fibre spectra, with a
diameter of $3$ arcsec. We have explored how our results change
  depending on the amount of the galactic light that those $3$ arcsec
  comprise. For this purpose, we have studied the behaviour of a
  subsample of galaxies with $R50\le 1.7$ arcsec, i.e. galaxies
with more than half of their light contained within the SDSS fibre spectra. This subsample exhibits similar tendencies
to the global ones, except that in this case there are galaxies with steep colour gradients that have their light highly concentrated. A subsample of galaxies with only
their cores measured within the SDSS spectra, with $R50>4.6$ arcsec, give a
better match to the global trends,
in the sense that the light concentration distributes in a
similar way to the total sample. We also find
similar trends for the  $D4000$ vs. $C$, when using the spectral index
derived by the MPA/JHU collaboration. Therefore, the tendencies of the
colour gradients with the SDSS age and metallicity indices can be
considered as representative of the tendencies with the global ages and
metallicities of galaxies without making any further consideration. 

We find that extreme internal colour variations tend
  to happen within galaxies that are blue, with low concentration
  indices and in the faint end of the sample distribution. Though,
  the difference between galaxies with
  extreme positive and negative colour gradients is small. This result
  implies that if the internal colour variation
  was mainly due to supernova winds or localised star formation
  regions, then the spatial probability for these phenomena to happen
  should be quite flat within galaxies. In their work on early-type
  galaxies \citet{tim10} suggested that for galaxies with a velocity
  dispersion below $140\, km/s$, metallicity gradients are tightly
  correlated with stellar mass, while above this threshold mergers are
  the main driver for setting the colour distribution within a
  galaxy. Thus, rather than a similar origin for blue and red cores, we could be looking to
  different phenomena: AGN or gas poor mergers vs. gas rich mergers,
  occurring with a similar probability. All this points out the
  difficulty on disentangling how very steep gradients were
  originated. Further work will be needed to understand the similar
  distributions of positive and negative colour gradients. 

\begin{figure*}
\begin{minipage}{0.32 \textwidth} 
{\epsfxsize=5.5truecm \epsfbox[35 26 568 476]{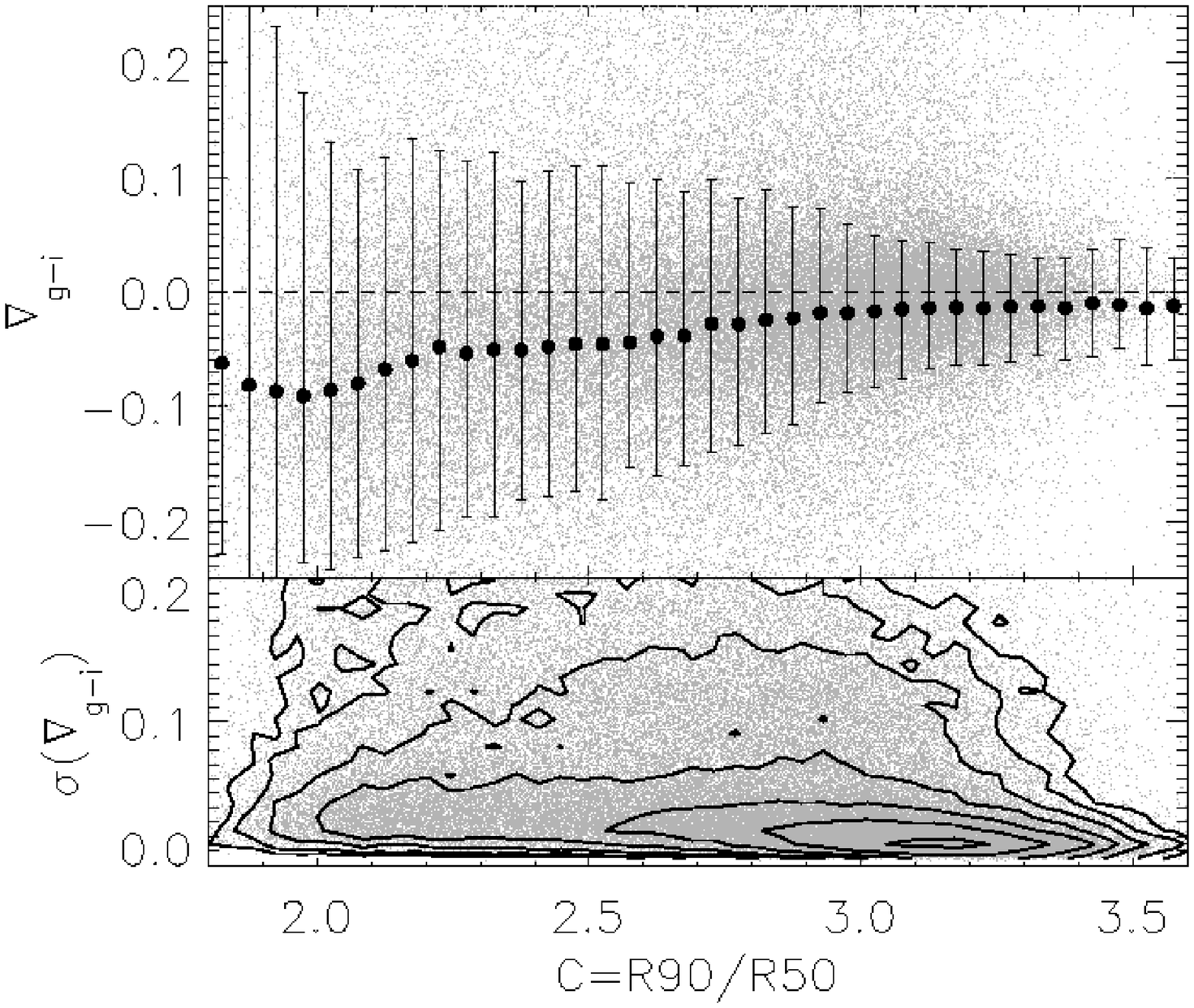}}
\end{minipage}
\   \
\hfill \begin{minipage}{0.32 \textwidth}
{\epsfxsize=5.5truecm  \epsfbox[35 26 568 476]{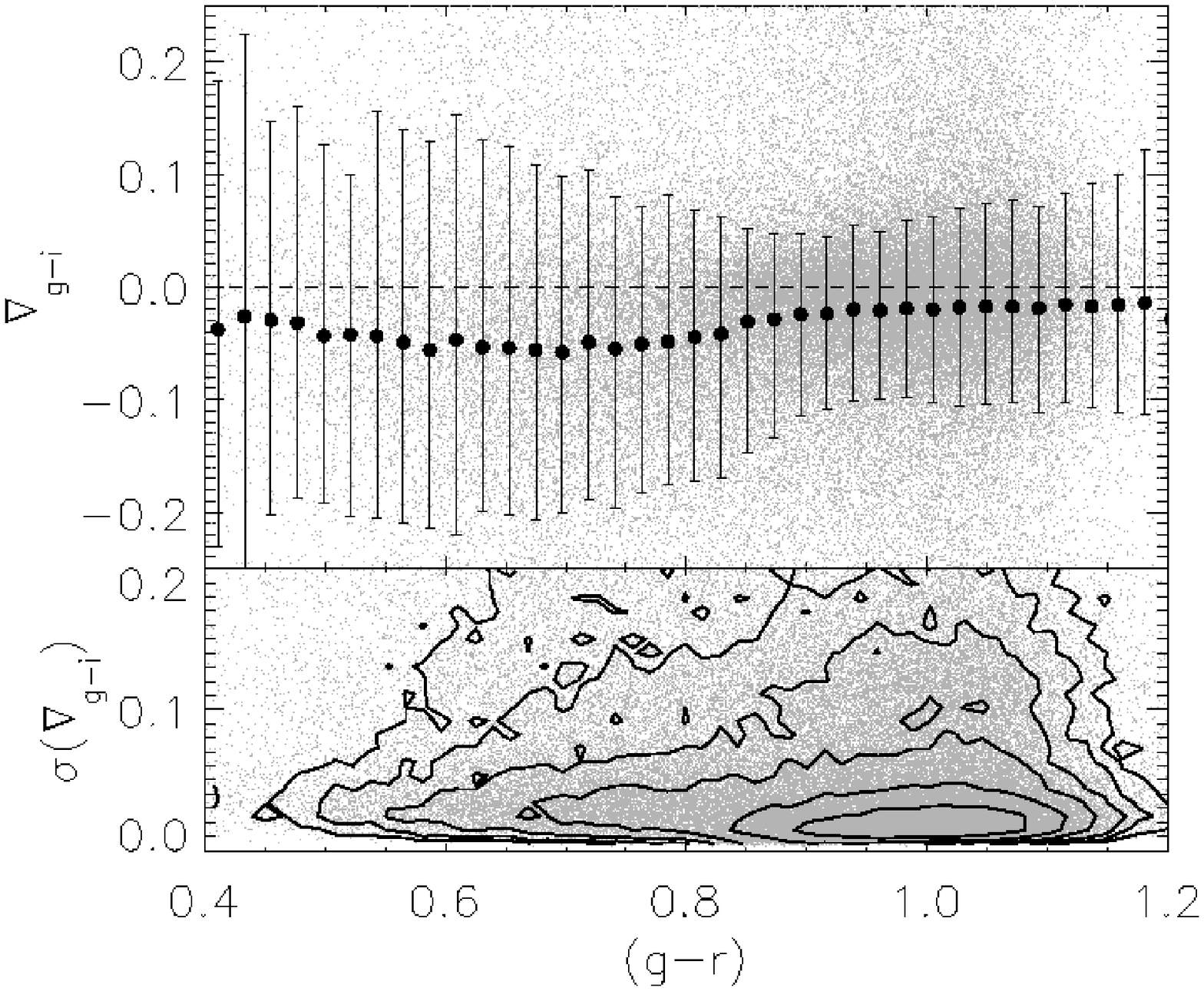}}
\end{minipage}
\   \
\hfill \begin{minipage}{0.32 \textwidth}
{\epsfxsize=5.5truecm  \epsfbox[35 26 568 476]{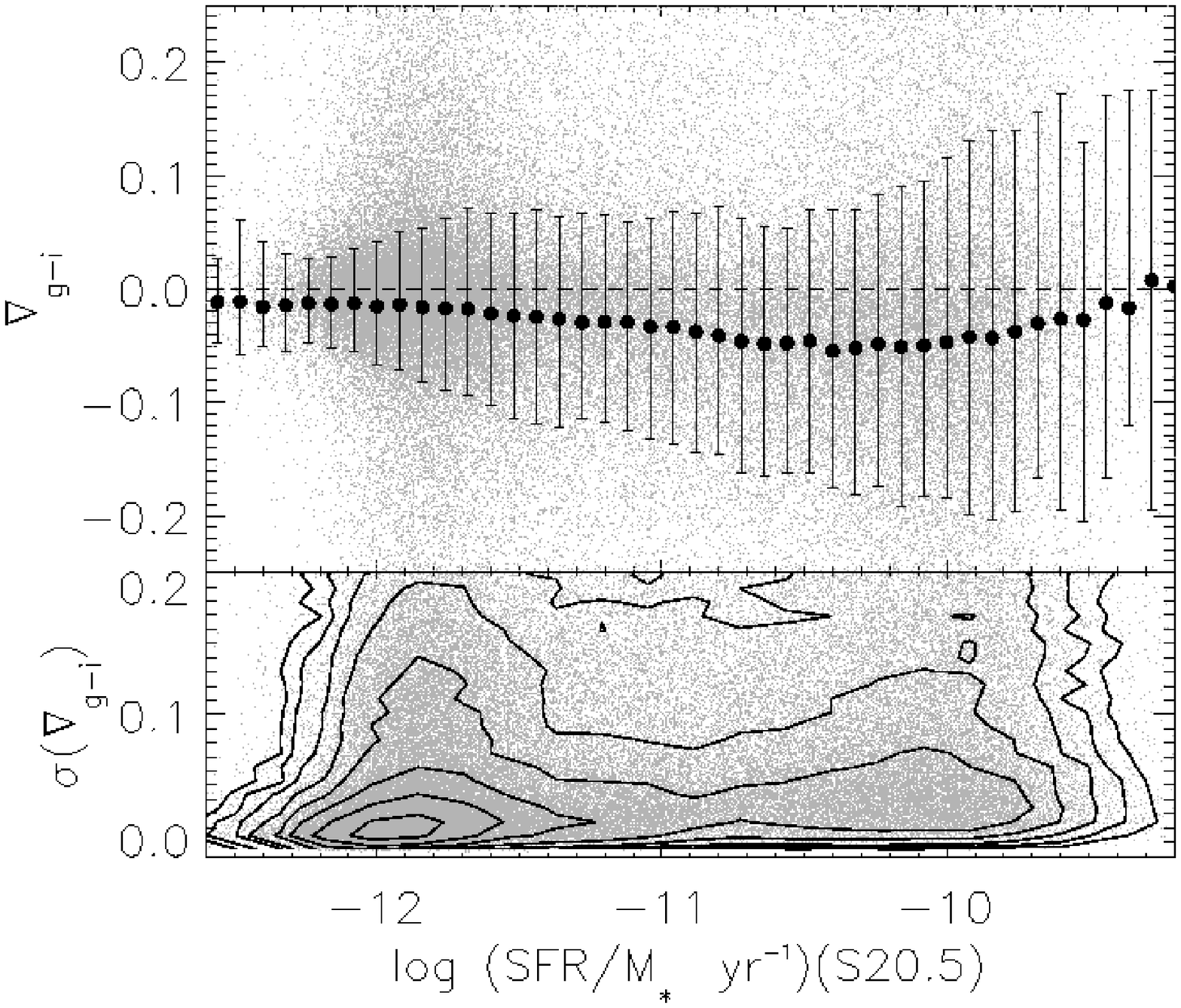}}
\end{minipage}
\caption{ Variation with the concentration index (left),
    $(g-r)$ colour (middle) and specific star formation (right) of the
    $(g-i)$ colour gradient, top
    panels, and its error, bottom panels, for galaxies within the
    {\it S20.5} sample. The panels showing colour gradients have
    superimposed median colour gradients values and their $1\sigma$
    dispersion range. The panels showing the variance of the colour gradients present density contours with increments of a factor of two.
}
\label{fig:el}
\end{figure*}

\subsection{Colour gradients within early and late type galaxies}\label{sec:types}

In this section we study the internal colour variation within
  early and late-type galaxies, using different parameters to classify
galaxies into one of those categories.

The left panel in Fig. \ref{fig:el} shows the variation of the $(g-i)$ colour gradient with the
concentration index, $C$,
for galaxies in the {\it S20.5} sample. As it can be seen in this plot, median
colour gradients tend to be flatter for galaxies with higher values of
the light concentration.  In fact, the same is seen for all the
studied colour gradients: $(g-i)$, $(g-r)$, $(r-z)$, $(r-i)$, $(i-z)$,
in the different luminosity-threshold samples considered here. This
was expected given the trends seen in the previous section. Due to the dispersion, this correlation is not statistically robust,
  with a Spearman coefficient of $0.19$. However, the increase in the
  dispersion of observed colour gradients is not correlated with a
  trend in the gradient errors and so, it is significant.

We find that early-type galaxies
tend to have flatter colour gradients than late-type galaxies. Most
spirals are classified as late-type and have an exponential light
profile. In simple terms, these galaxies are formed by a bulge and a disk
component. Bulges share features with early-type galaxies, being
dominated by an old stellar population. Meanwhile, disks contain, in
general, regions of star formation, providing an increase in the percentage of young stellar populations. These differences allow for greater
colour variations to happen in comparison with early-type galaxies,
whose characteristics are much more homogeneous.  

\citet{tortora10} also found a similar trend for colour gradients
with the S\'ersic index, which is tightly related to the
concentration index. These results also agree with the study by \citet{weinmann09}, were they found that galaxies with $C<3$ present steeper colour
gradients. 

Colour has also been widely used in the literature to split galaxies
into early and late-types. The middle panel in Fig. \ref{fig:el} shows
the variation of the $(g-i)$ colour gradient with the global $(g-r)$
colour of galaxies. In agreement with Fig. \ref{fig:color_gr} and
Fig. \ref{fig:color_ur}, it is clear in Fig. \ref{fig:el} that the
redder a galaxy is the flatter the colour gradient is expected to
be. The trend is similar to that seen for the concentration index:
early-type galaxies present flatter colour gradients. However, the
correlation with global colours is weaker than with the
concentration index, with Spearman index of $\sim 0.10$. This
could indicated that colour gradients depend more strongly on
morphological characteristics than on global colours. For example, if the
 colour gradient is set by the relative contribution of bulge and disc for late-type galaxies, one would expect that it would correlate more strongly with concentration index than with global colour.

For a given mass range, late-type galaxies are, in general, forming stars more efficiently
than early-types ones. The right panel in Fig. \ref{fig:el} presents
the variation of colour gradient with the specific star formation
derived by the MPA/JHU collaboration. With a Spearman index of
  $\sim 0.14$, the correlation between
  these two parameters is not statistically robust, likely due to the
  increase of the dispersion for those galaxies with higher specific
  star formation rates. The variance of the colour gradients does not
  determine the dispersion increase, which makes, again, the tendency significant. 

Assuming a constant star
formation, the inverse of the specific star formation, $M_*/SFR$, will
provide an estimation of the time it took a galaxy to build up its
observed stellar mass. Thus, higher values of specific star formation
will correspond to younger galaxies, or at least, with a larger
percentage of young stars. The found correlation between the specific
star formation and the colour gradient indicates that steeper
gradients appear for galaxies with higher percentages of young
stars. 

From Fig. \ref{fig:el} we can conclude that early-type galaxies
  have smaller internal colour variations compared to late-type
  galaxies.

\begin{figure}
{\epsfxsize=8.5truecm \epsfbox[68 17 546 473]{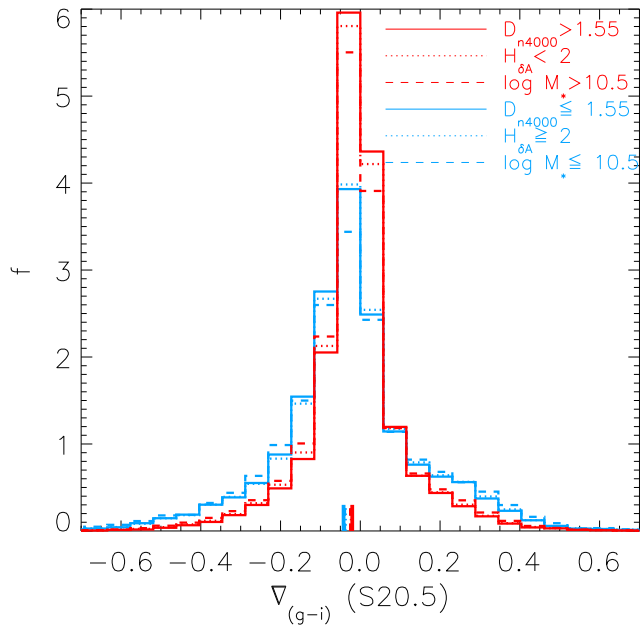}}
\caption{ Distribution of the $(g-i)$ colour gradients within galaxies
    from {\it S20.5} sample split
    into early (red lines) and late (blue lines)
    types using different parameters: $4000$ {\AA} break, solid lines, 
    $H_{\delta A}$ Balmer index, dotted lines, stellar mass, dashed
    lines. Median values are shown as short, thick lines of the corresponding type. Histograms are normalised to give unit area underneath.}
\label{fig:el_dis}
\end{figure}

We have also compared the shape of the distribution of colour
 gradients within early and late-type galaxies, selected
accordingly to either concentration index, global colour or specific
star formation, finding that early-type colour gradients have narrower
distributions than late-types. As can be seen in Fig. \ref{fig:el_dis} this is also true when separating
galaxies into early and late-types using other parameters. 

Fig. \ref{fig:el_dis} shows the $(g-i)$ colour gradient distribution for
galaxies from the {\it S20.5} sample
separated into early and late-types following the \citet{kauffmann_tot:03}
work, by
using the following parameters derived by the MPA/JHU collaboration: the $D_n4000$ index, sampling the spectral $4000${\AA}
break, the Balmer index $H_{\delta A}$ and the logarithmic stellar
mass. Mass does not clearly separate galaxies into early and
  late-types, but the characteristics of galaxies with masses below or
above $10^{10.5}M_{\odot}$ are rather different \citep{kauffmann_tot:03}. Despite median values being quite close for early and late-types
galaxies, it is apparent that late-type galaxies present a wider
distribution with more extended tails than for the case of the
early-type galaxies distribution. 

As it is found here, \citet{tortora10}  also reported a larger
dispersion for colour gradients among late-type galaxies. \citet{suh10} found that although most of the early-type
galaxies have quite flat colour gradients, there is $\sim 30$\% with
residual star formation that produces steeper colour gradients, in
consonance with the distributions reported here.

Following \S \ref{sec:extreme}, we have also studied here the relation between global galactic colours and
  the existence of extreme colour gradients within galaxies with low
  and high concentration indices. We find again that steep colour
  gradients appear within bluer late-type galaxies. However, in the
  case of early-type galaxies the
  global galactic colour does not appear to be correlated with the
  existence of
  extreme colour gradients.

When defining colour gradients as the variation of colour with
  the logarithm of the normalised
radius, we obtain that early-type galaxies ($C>2.6$) have a median $(g-i)$
colour gradient of $-0.13$ for both the {\it S19} and {\it S20.5}
samples, while late-type galaxies have median values of $-0.27$ and $-0.32$, for
those two samples respectively. The quoted values for early-type
galaxies are smaller than those found by \citet{lb10}. In the study from
\citet{lb10}, they apply two cuts in order to minimise the
contamination of the early-type galaxy sample. Since we have only
separated galaxies on basis of their concentration index values, we
could have a larger fraction of late-type contaminants, providing
steeper colour gradients than those reported by  \citet{lb10}.

In summary, the probability for finding strong internal colour variations within
early-type galaxies is lower than for late-type galaxies.

\subsubsection{Early and late type galaxies based on their colour gradients}

\begin{figure}
{\epsfxsize=8.5truecm \epsfbox[61 23 684 490]{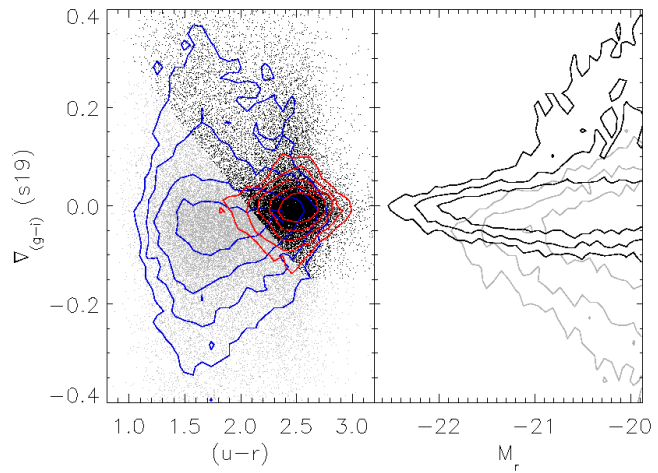}}
\caption{ Both panels show data from galaxies within the {\it S19} sample. Galaxies have been separated into
  early, in black, and late-type galaxies, in grey, following
  \citet{park05} criterion. Contours represent a factor of two change in
  density. {\it Left:} $(u-r)$ colour vs. $(g-i)$ colour
  gradient. Blue contours show the density for galaxies with $C \le
  2.6$, those in red correspond to galaxies with $C>3$. {\it Right:} Absolute magnitude in r band vs. $(g-r)$ colour gradients. 
}
\label{fig:choi}
\end{figure}

\citet{choi07} have studied the behaviour of colour gradients with
absolute magnitude for early and late-type galaxies, separated by
their position in the colour gradient versus colour parameter space as
defined by \citet{park05}.  In order to compare our results with the
\citet{choi07} study we have also separated galaxies into early and
late-type according to the \citet{park05} method. Our definition
of colour gradients is different since they just compare the light
outside and inside half the Petrosian radius of
galaxies. Moreover, they use magnitudes K-corrected to redshift
$0.1$. We have not applied any corrections to take
into account these differences, we will thus compare tendencies
qualitatively, rather than actual values.

The left panel in Fig. \ref{fig:choi} shows the $(u-r)$ colour versus
the $(g-i)$ colour gradient, using black/grey dots for those
early/late-type galaxies according to the \citet{park05} criterion. In the
same panel we have superimposed density contours for galaxies with
different light concentrations: $C \le 2.6$ and $C>3$. We can see
there that the \citet{park05} classification scheme agrees quite well
with separations based on the concentration index.

The right panel in Fig. \ref{fig:choi} shows the $(g-i)$ colour
gradients as a function of absolute magnitude. Fig. \ref{fig:choi}
shows that those galaxies classified as early-types have a very
constant median colour gradient with absolute magnitude. The same
  behaviour is reported by \citet{choi07}  and \citet{lb05}, though other studies find some trends \citep[see e.g.][]{lb10}.
Early-type galaxies within the {\it S19} sample, especially those
fainter than $M_{r}\sim -21$, present values with increasing
dispersion towards positive values, i.e. towards having bluer
cores. This could be related to the branch of early-type
galaxies with residual star formation, as the E+A galaxies\footnote{E+A galaxies
  are dominated by an old stellar population but their spectra show
  signatures of the occurrence of minor star formation bursts that have recently ceased. See
  \citet{yamauchi05} for a study of colour gradients in E+A
  galaxies.}, selected by \citet{park05} method. Indeed, in their work they
showed that one of the advantages of their way of separating galaxies
is that E+A galaxies are classified as early-types, although they are
bluer and less concentrated than most early-type galaxies. In
  fact, this increase in dispersion towards bluer cores within
  early-types is the only difference found in the colour gradient
  vs. absolute magnitude parameter space when using the concentration index to
separate galaxies into early and late-types.

Unlike global galactic colours, the internal colour variation do not
relate clearly to the luminosity of galaxies, though  median colour
gradients tend to be steeper at brighter absolute magnitude bins. For fainter magnitudes than the ones studied here, \citet{choi07} observed that the colour gradients of late-type galaxies increase with magnitude. Nevertheless, within the magnitude range studied here the median colour gradients reported by \citet{choi07} also stay quite flat with magnitude. \citet{bakos08} studied the internal colour variation within $85$ spiral galaxies  finding that it is almost constant for a given luminosity. 

In comparison to early-types, late-type galaxies have steeper colour
gradients and present a larger dispersion that increases for fainter
magnitudes. These results are in agreement with \citet{choi07} and in
consonance with our previous findings.

\subsubsection{Velocity dispersion and stellar mass}\label{sec:mass}

\begin{figure}
{\epsfxsize=8.5truecm \epsfbox[23 10 556 462]{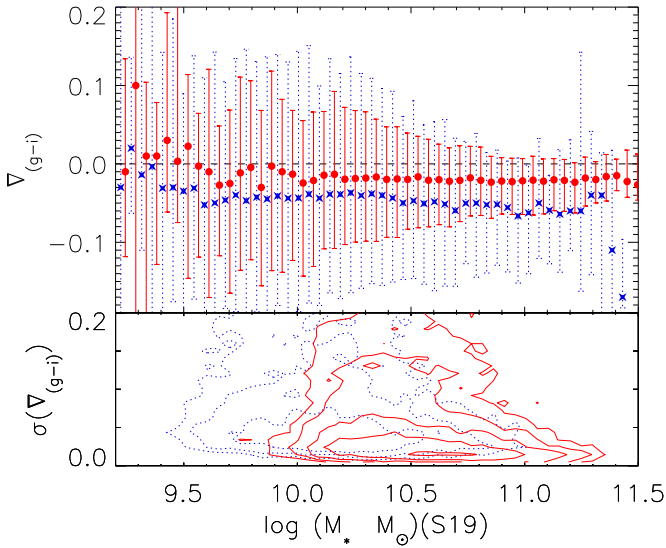}}
\caption{ Variation with stellar mass of the median $(g-i)$ colour gradient
  for early, circles, and late-type galaxies, stars, within
  sample {\it S19}. Error bars corresponds to the $1\sigma$ range in
  dispersion. The bottom panel shows the variance of colour gradients
  within early, solid contours, and late-type galaxies, dotted
  contours. Plotted contours increase by a factor of two in density.}
\label{fig:lgm}
\end{figure}

Due to the complexity for obtaining the velocity dispersion, the SDSS
provides measurements only for galaxies with a clean spectrum typical
of an early-type at $z < 0.4$. The subsample of galaxies with a
reliable measurement of the velocity dispersion presents slightly flatter colour gradients with larger velocity
dispersions. Such tendency is not statistically robust for the $(g-i)$
colour gradient, with an Spearman index of $\sim 0.12$, due to the
increase in dispersion. Though, as for the parameters studied before, the
distribution of errors
alone cannot explain the change in dispersion, which makes the
tendency significant. The same tendency has been found in the
  study by \citet{roche10} on early-type galaxies.

Stellar mass and velocity dispersion are tightly correlated. The trend
seen for the velocity dispersion is supported by the correlation that we find
between the $(g-i)$ colour gradient and the stellar mass derived by
the MPA/JHU collaboration. For a similar range in both masses and luminosity,
our result for the most massive galaxies to present flatter colour
gradients agrees with previous studies
\citep{weinmann09,suh10,tortora10}. 

Fig. \ref{fig:lgm} shows the variation with stellar mass of the
  median $(g-i)$ colour gradient for early, with $C>2.6$, and
  late-type galaxies, with $C\le 2.6$. The median of the colour gradient distributions for both type
  of galaxies remain quite flat for masses above $M\ge
  10^{9.5}M_{\odot}$. Less massive galaxies tend to present flatter or even
  positive colour gradients, although the dispersion is much larger.

Our sample of galaxies contain rather massive galaxies, with $M\ge 10^{9}
M_{\odot}$. For this range of masses there is a wide dispersion for
the star formation rates \citep{brinchmann:04}, thus it is difficult
to connect this result with the different mechanisms that are able to
produce colour gradients. Splitting our sample into early and late types we find that only those
early-type galaxies with $M_*<10^{9.5}M_{\odot}$ within the {\it S19}
sample present a median positive colour gradient. \cite{spolaor09}
have studied the dependency of metallicity gradients with stellar mass
finding them to be tightly related if $M_* \le 3.5\times 10^{10}\,
M_{\odot}$, as predicted by pure monolithic collapse models
\citep[e.g.][]{chiosi:02}, while more massive galaxies do not appear
to show this relation, as other studies previously proved
\citep{michard:05,sanchez-blazquez,tortora10}. For those galaxies in the lowest
stellar mass range, the metallicity gradient becomes shallower with
decreasing mass, which could translate into age gradients having a
dominant effect on the colour gradients. This will provide the
possibility to invert the colour gradients. In general, those early-type
galaxies  at $z=0$ with the lowest mass could present higher star
formation rates than more massive ones \citep{delucia06}, which in turn could provide the existence of
steeper positive colour gradients (bluer cores). We find a similar
result for late-type galaxies, in the sense that only those galaxies
in the lowest stellar mass bins present null or positive median colour
gradients values.

The tendency found for early-type
galaxies in Fig. \ref{fig:lgm} agrees with the studies from
\citet{suh10} and \citet{tortora10}, supporting that bluer cored
early-type galaxies only appear in the low mass bins.

In the case of late-type galaxies, localised regions of star formation
can affect the colour gradients but dust and stellar migration can also
have a significant impact on them \citep{sb09,vlajic09}. Stellar
migration happens immediately after the formation of the wave density
that leads to the formation of spiral arms, but also after the
formation of bars or disk instabilities that can be triggered by tidal
forces. Stellar migration is strongest for the most massive galaxies
and tends to wipe out initial metallicity gradients, and, therefore,
colour ones. This effect rises age gradients opposed to the
metallicity ones, though the dispersion for these
increase with mass \citep{tim10}. The combined effect of age and
metallicity gradients for massive galaxies can explain why we find the
colour gradient to stay quite flat for late-type galaxies of different
masses. \citet{tortora10} found a clear steepening with
mass for the colour gradient within late-type galaxies. As mention before, we find that the least massive galaxies in our sample present flatter median colour gradients, i.e., we find a very slight trend in the same direction as the one reported by \citet{tortora10}. However, the slopes of these trends are different. The origin of this difference is unclear but it could be related with the different definitions of colour gradient or the sample selection.

\subsection{Colour gradients, star formation and nuclear activity}\label{sec:agn}

\begin{figure}
{\epsfxsize=8.5truecm \epsfbox[53 44 507 488]{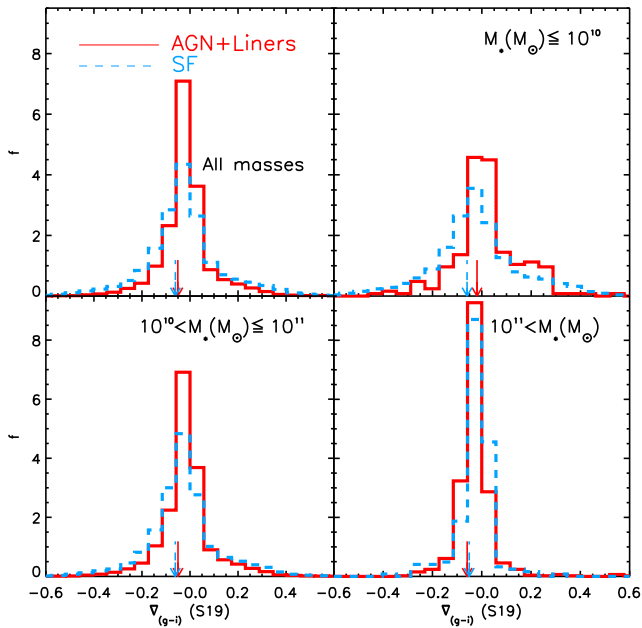}}
\caption{ Distribution of the $(g-i)$ colour gradients within galaxies
    from the {\it S19} sample, separated into AGN plus LINERS (solid
    line) and star forming galaxies (dashed line) following
    \citet{brinchmann:04}. The top left panel shows the distribution
    for all the galaxies in the sample, while the rest show the
    distribution within the mass range specified in the legend. Arrows show median values. Histograms are normalised to give unit area underneath.}
\label{fig:agn}
\end{figure}

Beside stellar winds and migrations, the nuclear activity could also
be responsible for the existence of anomalous internal colour
variations within galaxies. In fact, \citet{menanteau:05} found one blue
nucleated spheroid with star formation associated with the presence of
an AGN. 

Here we try to disentangle the effects of star formation and
  nuclear activity as responsible for setting colour gradients. We
use the classification done by the MPA/JHU collaboration for the SDSS
DR7 based on the one from \citet{brinchmann:04}, which makes use of the location of a galaxy in the parameter space
  $log([OIII]5007\diagup H_{\beta})$ versus $log([NII]6585\diagup
  H_{\alpha})$, the so called BPT diagram \citep{baldwin:81}, taking
  into account the signal to noise ($S/N$) of the spectral lines. In
this parameter space it is possible to separate star forming galaxies
from those with nuclear activity, though there is a region were
galaxies are considered as composites. Due to the requirement of a
minimum $S/N$ value for the lines, not all galaxies can be classified
following this scheme.

We have explored the locus occupied in the BTP diagram by those galaxies with very steep
colour gradients, finding that they distribute following the mean trend.

Fig. \ref{fig:agn} presents the distribution of the $(g-i)$ colour
gradients within galaxies from the {\it S19} sample, separated into
galaxies with nuclear activity and star forming galaxies. The left upper panel of this figure presents the whole sample, showing that steep
colour gradients are more likely to appear within star
forming galaxies rather than those with nuclear activity. In fact, there is a minute increase, $\sim 2$\%,  in the percentage of galaxies classified as star
forming among those with steep colour gradients, compared with
the mean trend. This last tendency goes in the direction expected
from the above discussions.

The studied early-type galaxies present a
higher percentage of galaxies with nuclear activity, independently of
their colour gradients. Thus, the top left panel in Fig. \ref{fig:agn} could be driven by the different distribution of colour
gradients for early and late-type galaxies seen in \S\ref{sec:types}. 

The remaining panels in Fig. \ref{fig:agn} explore the colour gradient distributions for
galaxies within three mass ranges. \citet{agn} found that different
levels of nuclear activity appear in galaxies with different
masses. Galaxies with masses below $10^{10}M_{\odot}$ could present a
weak AGN, but not a strong one. In this mass range we observe that
those galaxies from the {\it S19} sample that are hosting an AGNs present
a higher percentage of bluer cores than those forming stars. The other
studied samples of galaxies do not present this difference, but yet,
they contain fewer low mass galaxies. The observed trend could be
  explained if those weak AGN were triggering star formation localised
  only in the central region of their host, or, more unlikely, if the
  AGN within these low mass galaxies was driving dust out of the central galactic region,
lowering its internal extinction, providing bluer cores.

Galaxies more massive than $10^{10}M_{\odot}$, host most of the AGN
activity. As expected from \S\ref{sec:mass}, the colour gradient
distribution for the most massive galaxies is quite narrow, with few
galaxies presenting steep colour gradients. The median colour
gradients for massive galaxies remain quite constant and comparable to
the global median.

Our results suggest that nuclear activity is a marginal driver for
creating steep colour gradients, and that it is overcome by the dependency
between colour gradients and morphological type, at least for those galaxies with masses above $10^{10}M_{\odot}$.

\subsection{Colour gradient change with redshift}

\begin{figure}
{\epsfxsize=8.5truecm \epsfbox[23 20 574 463]{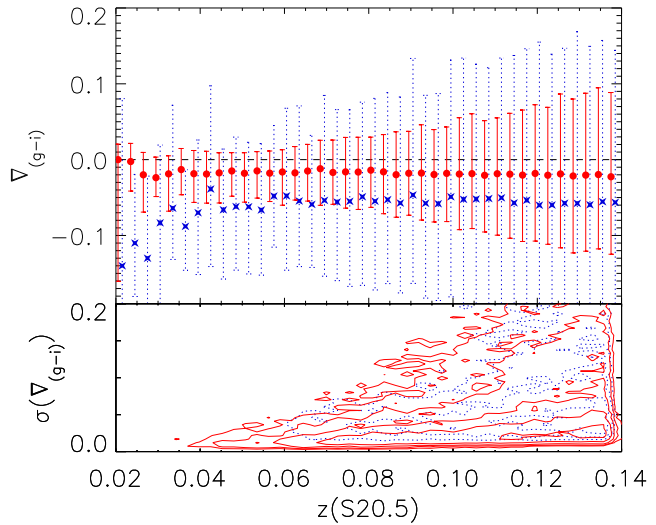}}
\caption{ Similar to Fig. \ref{fig:lgm} for the variation with redshift of
  the $(g-i)$ colour gradients within
galaxies in sample {\it S20.5}.}
\label{fig:z}
\end{figure}

Comparing the two panels in both Fig. \ref{fig:color_gr} and
Fig. \ref{fig:color_ur} for the samples under study, we have observed
that galaxies with steep colour gradients tend to be redder when
moving to samples including galaxies at higher redshifts. The global
sample also gets redder when going from sample {\it S19} to {\it
  S20.5}, but the variation is smaller. We have observed that colour
gradients do not depend strongly on absolute magnitude, except for an
increase in the scatter at the faintest end. Thus, this result could
suggest a stronger evolution for those galaxies with steeper colour
gradients in the recent past.

Fig. \ref{fig:z} shows the variation of the $(g-i)$ colour gradient
with redshift for early and late-type galaxies within the {\it S20.5}
sample. For both types of galaxies the median colour gradients
  stay constant with the redshift, while the dispersion and errors
  increase. We find that for a fixed magnitude the dispersion in the
  colour gradient as a function of redshift is highly reduced. The
  opposite is not true. This is in agreement with \citet{bakos08}.

We observe that it is more probable to find a galaxy with a steeper colour gradient at higher redshift. However, this trend of the dispersion could be driven by the errors distribution.

We have explored visually 48 early-type galaxies randomly selected among those
with positive gradients, outside the $1\sigma$ dispersion range, and within the
median value to explore the possibility of those early-type galaxies with steeper
colour gradients to be more affected by artifacts. We find the same
amount of galaxies either close to stars or other galaxies among both
groups of galaxies. Thus, the observed tendency does not appear to be due to a higher presence of artifacts among those early-types with steeper colour gradients.

In agreement with our study, \citet{lb03} found that colour
  gradients within spheroids are weakly dependent on the redshift, up to
  $z\sim 0.65$. Colour gradients within the disks studied by
  \citet{lb03} change
  appreciably from $z\sim 0$ to $z\sim 0.65$. However, this change is
  not so clear within the redshift limits used in our study. 

The fact that median colour gradients do not appear to evolve at
$z<0.17$, points out that most of the star formation activity occurred previously.


\section{Summary and conclusions}\label{sec:conclusions}

In this paper we have studied the internal colour variation within
galaxies from the SDSS DR7, with both clean photometry and
spectroscopic measurements. We have obtained the global
(K+e)-correction by using SED templates from {\sc pegase} and applied
them to each galaxy in order to divide our initial sample into
luminosity-threshold ones, with redshifts up to $z=0.17$. These
samples are the largest used for studying the internal colour
variation in galaxies.

The internal colour variation has been quantified through the slope of
the relation between radial colour and the galactic radius normalised
to the one containing half the total galactic light,
$R50$. Colour gradients have been obtained from the SDSS measurements
of surface brightness circularly averaged in concentric
annuli. Observational errors have been taken into account using
  a Monte Carlo method in the colour gradient calculation. This
approach is independent of previous ones in the literature, therefore
the agreement with earlier works underscores the robustness of the
colour gradient as a useful parameter for studying particular aspects
of galactic evolution.

In agreement with previous studies, we have found that colour
gradients distribute around median values that are negative but close
to zero, i.e. galaxies, independently of their morphological type,
tend to have slightly redder cores than their external parts. 

We have explored some of the possible systematics that could affect
the measurement of the colour gradients. In particular we have
explored the possible dilution due to the fact that we start our
  calculation of the colour gradient from a circularly averaged
  surface brightness profile. We observe that very elongated galaxies
tend to have steep colour gradients, the opposite trend to the one
  expected if they were diluted. According to \citet{choi07}, steep
colour gradients within elongated galaxies are due to their internal
dimming. In the calculation of the colour gradients we have left out
the innermost annuli, since they are the ones most affected by
seeing. We have explored the effect of obtaining colour gradients
using different number of annuli, finding that for the range in which
we work our results on colour gradients are robust. We find that
galaxies with larger half light radii tend to have steeper colour
gradients, being an intrinsic characteristic of galaxies and not an
artifact.

We have selected a subsample of close pairs of galaxies that appear to
be interacting. The sample is small, far less than $1$\% of the total
sample, and it represents an inferior limit of such close pairs of
galaxies with respect to what the total percentage that might be found
when considering the whole photometric SDSS sample. Despite the small
number of galaxies in this sample, we find a larger proportion of
galaxies with bluer cores here than in the whole population sample. As
close pairs are more likely to be interacting and normally these
interactions are associated with an enhancement of star formation,
this supports the connection between steep colour gradients and recent
bursts of star formation.

We find that the colour gradient distribution does not follow a
Gaussian due to an excess of galaxies in the tails that cannot be
explained alone by the errors distribution. A double-Gaussian fits
better the colour distribution, pointing out a possible
contribution of two distinctively different galaxy classes. Separating
the initial sample of galaxies into early and late-types we find that
the distribution for colour gradients within late-type galaxies is
more extended than that for early-types. Nevertheless, the
distribution of early-type galaxies is still broader than a
Gaussian. We find this excess of steep colour gradients to be likely
related to age gradients, AGN activity or an unusual dust content.

The observed distribution of colour gradients can be understood
as a consequence of entangled effects. Gradients produced in mergers
depend on the time scales that interplay in their creation and the
mixing of elements that can contribute to their
destruction. Galaxies observed at different epochs after a
gravitational interaction can present colour gradients evolving with
time. Nevertheless, even when observing galaxies at a similar time
after a merger one expects to find a distribution of colour gradients
and not a single value, since the metallicity enrichment appear to be
highly non-uniform, due to the multi-phase nature of the inter
galactic medium \citep{springel03}. Besides, \citet{kobayashi:04}
simulated the formation of early-type galaxies through different kind
of mergers, finding that the metallicity gradients (the main origin of
colour gradients) strongly depend on the ratio of stellar masses and
gas content between the progenitors. Moreover, even if early-type
galaxies were formed in a "monolithic collapse", for a given stellar
mass there would be certain scatter in star formation efficiencies
which provides a scatter in colour gradients
\citep[e.g][]{pipino10}. For late-type galaxies, the colour
  gradient dispersion is further enhanced due to the dust content and the bulge to disk relation
  depending on the observing inclination angle. The complicated patterns of stars migrations that follow the
spiral arm formation can also enhance, at least temporally, this dispersion \citep{sb09}.

Galaxies with steep gradients share some global properties
independently of presenting bluer or redder cores. Nevertheless,
redder cores happen, on average, within galaxies redder than those
with bluer cores. We have also observed that galaxies with high
concentration values do not present steep colour gradients. It is
known that mergers increase the galactic size and that galaxies with
highly concentrated light have features indicating either a passive
evolution in their recent past or the lack of gas, ruling out the
possible existence of great internal colour variations. Thus, both
results suggest the existence of a relation between colour gradients
and the recent formation history of galaxies, in the sense that very
steep colour gradients are likely related to recent star bursts.

We have found that galaxies with steeper colour gradients are more likely
to be late-types according to either their structural parameters or to
their specific star formation, i.e., a higher presence of young
stars. It is interesting to note that we have found that colour
gradients are less tightly related to global colours than the
concentration index. \citet{park05} found out that the parameter space
colour-colour gradient could be very useful to select galaxies
matching their optical morphology at low redshifts. Our study
indirectly supports theirs, suggesting that colour gradients origins
are more related to the recent formation history of galaxies, the kind
of gravitational interactions recently suffered, than with the global
galactic colours. Different studies \citep[see e.g.][]{zoo:blue} are
finding spheroidal galaxies with blue colours and red spirals. This
proves the usefulness of a parameter like the colour gradient that is
sampling something that cannot be seized through the global colour.

We have found that the possible correlation between colour gradients
and nuclear activity is overcome by the stronger relation with
galactic type, since nuclear activity appears more common among
early-types in our sample of galaxies.

At intermediate redshift, $0.37<z<0.83$, \citet{menanteau:01} found
that $30$\% of field spheroids (classified based on their
morphological parameters) show an unusual internal colour
variation. \citet{menanteau:01} associated this large colour variation
with either star formation or AGN activity, in consonance with our results for a closer sample of galaxies. 

Within the redshift range studied here, the colour gradients remain
rather constant, implying that the studied galaxies had
their last residual bursts of star formation at earlier times. 

The results presented in this work agree with previous studies,
though, here we are more dominated by the scatter due to the larger
sample explored. Nevertheless, our results are very robust since they
hold for the four studied luminosity-threshold samples. 

Colour gradients are normally defined in the literature as the colour variation as a function of the logarithm of the radius. Here we have used the variation of colour as a function of the radius itself. 
Using the former definition we have checked that we
  obtain qualitatively similar trends for colour gradients, except
  that their distribution covers a broader dynamical
  range. For the {\it S20.5} sample, the median values of these distributions are $-0.187$ and
  $-0.277$, respectively for the $(r-z)$ and $(g-i)$ colour
gradients. In our calculation of
colour gradients we assume the observational errors in both colours
and radii to be normally distributed. When using the logarithm of
normalised radius, the error distribution deviates from a
Gaussian. Thus, in order to make an accurate comparison we should modify the way we are including the observational errors in our
calculation. This is likely to have a fundamental effect on the
estimated errors, while colour gradients would remain practically unchanged. Leaving apart this issue, when defining the colour
gradient as colour variation as a function of the logarithm of the radius we find qualitatively similar
tendencies to the ones previously reported. The correlations
appear stronger when defining colour gradients in this way, due to the
enhanced dynamical range.

The fact that steep colour gradients are linked with
a higher presence of young stars could provide an alternative method
for exploring the interactions between galaxies. A thorough spectral
analysis of those galaxies which present extreme colour variations
could help us to better understand the mechanisms that generate either bluer or redder cores within galaxies.

\section*{ACKNOWLEDGEMENTS}
{
      We thank J. Lucey, I. Ribas and R. Smith for their helpful
      comments. Thanks to C. Baugh for reading the manuscript. VGP is supported by a Science and Technology Facilities 
Council rolling grant. VGP and FJC acknowledge support from Spanish
Science Ministry AYA2009-13936, Consolider-Ingenio CSD2007-00060 and
project 2009SGR1398 from Generalitat de Catalunya.

}
\vspace{-0.7cm}

\bibliographystyle{mn2e} 

\begin{thebibliography}{}

\bibitem[\protect\citeauthoryear{{Abazajian} et~al.}{2009}]{dr7}
{Abazajian} K. {et~al.},  2009, \apjs, 182, 543

\bibitem[\protect\citeauthoryear{{Bakos}, {Trujillo} \& {Pohlen}}{{Bakos}
  et~al.}{2008}]{bakos08}
{Bakos} J.,  {Trujillo} I.,    {Pohlen} M.,  2008, \apjl, 683, L103

\bibitem[\protect\citeauthoryear{{Baldwin}, {Phillips} \&
  {Terlevich}}{{Baldwin} et~al.}{1981}]{baldwin:81}
{Baldwin} J.~A.,  {Phillips} M.~M.,    {Terlevich} R.,  1981, PASP, 93, 5

\bibitem[\protect\citeauthoryear{{Begelman}, {de Kool} and {Sikora}}{{Begelman}, {de Kool} and {Sikora}}{1991}]{begelman91}
{Begelman} M., {de Kool} M. and {Sikora} M., 1991, \apj, 382, 416B


\bibitem[\protect\citeauthoryear{{Blanton} et~al.}{2003}]{blanton:03}
{Blanton} M.~R. {et~al.},  2003, APJ, 594, 186


\bibitem[\protect\citeauthoryear{{Brinchmann}, {Charlot}, {White}, {Tremonti},
  {Kauffmann}, {Heckman} \& {Brinkmann}}{{Brinchmann}
  et~al.}{2004}]{brinchmann:04}
{Brinchmann} J.,  {Charlot} S.,  {White} S.~D.~M.,  {Tremonti} C.,  {Kauffmann}
  G.,  {Heckman} T.,    {Brinkmann} J.,  2004, MNRAS, 351, 1151

\bibitem[\protect\citeauthoryear{{Bruzual} \& {Charlot}}{{Bruzual} \& {Charlot}}{1993}]{bruzual93}
{Bruzual A.} G. and {Charlot} S., 1993, \apj, 405, 538

\bibitem[\protect\citeauthoryear{{Charlot}, {Worthey} \& {Bressan}}{{Charlot}, {Worthey} \& {Bressan}}{1996}]{charlot96}
{Charlot} S., {Worthey} G. \& {Bressan} A., 1996, \apj, 457, 625

\bibitem[\protect\citeauthoryear{{Chiosi} \& {Carraro}}{{Chiosi} \&
  {Carraro}}{2002}]{chiosi:02}
{Chiosi} C.,  {Carraro} G.,  2002, \mnras, 335, 335

\bibitem[\protect\citeauthoryear{{Choi}, {Park} \& {Vogeley}}{{Choi}
  et~al.}{2007}]{choi07}
{Choi} Y.-Y.,  {Park} C.,  {Vogeley} M.~S.,  2007, \apj, 658, 884


\bibitem[\protect\citeauthoryear{{Clemens} et~al.}{{Clemens} et~al.}{2009}]{clemens09}
{Clemens} M.~S., {Bressan} A., {Nikolic} B., {Rampazzo} R., 2009,
\mnras, 392, L35

\bibitem[\protect\citeauthoryear{{Cooper} et~al.}{{Cooper} et~al.}{2009}]{acooper09}
{Cooper} A.~P.  {et~al.}, 2009, ArXiv e-prints

\bibitem[\protect\citeauthoryear{{de Boor}}{{de Boor}}{1978}]{deboor:78}
{de Boor} C.,  1978, {A practical guide to splines}.
Applied Mathematical Sciences, New York: Springer, 1978

\bibitem[\protect\citeauthoryear{{De Lucia}, {Springel}, {White}, {Croton} \&
  {Kauffmann}}{{De Lucia} et~al.}{2006}]{delucia06}
{De Lucia} G.,  {Springel} V.,  {White} S.~D.~M.,  {Croton} D.,    {Kauffmann}
  G.,  2006, \mnras, 366, 499

\bibitem[\protect\citeauthoryear{{Driver} et~al.}{2006}]{driver06}
{Driver} S.~P.,  {et~al.} 2006, \mnras, 368, 414

\bibitem[\protect\citeauthoryear{{Ferreras}, {Lisker}, {Pasquali} \&
  {Kaviraj}}{{Ferreras} et~al.}{2009}]{ferreras09}
{Ferreras} I.,  {Lisker} T.,  {Pasquali} A.,    {Kaviraj} S.,  2009, \mnras,
  395, 554

\bibitem[\protect\citeauthoryear{{Fioc} \& {Rocca-Volmerange}}{{Fioc} \&
  {Rocca-Volmerange}}{1997}]{fioc:97}
{Fioc} M.,  {Rocca-Volmerange} B.,  1997, A\&A, 326, 950

\bibitem[\protect\citeauthoryear{{Fukugita}, {Shimasaku} \&
  {Ichikawa}}{{Fukugita} et~al.}{1995}]{fukugita:95}
{Fukugita} M.,  {Shimasaku} K.,    {Ichikawa} T.,  1995, PASP, 107, 945

\bibitem[\protect\citeauthoryear{{Hamilton}}{{Hamilton}}{1985}]{hamilton85}
{Hamilton} D.,  1985, \apj, 297, 371

\bibitem[\protect\citeauthoryear{{Hinkley} \& {Im}}{{Hinkley} \&
  {Im}}{2001}]{hinkley:01}
{Hinkley} S.,  {Im} M.,  2001, \apjl, 560, L41

\bibitem[\protect\citeauthoryear{{Kauffmann}}{{Kauffmann}  et~al.}{2003a}]{kauffmann_mass:03}
{Kauffmann} G. et~al.,  2003a, MNRAS, 341, 33

\bibitem[\protect\citeauthoryear{{Kauffmann}}{{Kauffmann}  et~al.}{2003b}]{kauffmann_tot:03}
{Kauffmann} G. et~al.,  2003b, MNRAS, 341, 54

\bibitem[\protect\citeauthoryear{{Kauffmann}}{{Kauffmann}  et~al.}{2003c}]{agn}
{Kauffmann} G. et~al.,  2003c, MNRAS, 346, 1055

\bibitem[\protect\citeauthoryear{{Kobayashi}}{{Kobayashi}}{2004}]{kobayashi:04}
{Kobayashi} C.,  2004, MNRAS, 347, 740

\bibitem[\protect\citeauthoryear{{Ko} \& {Im}}{{Ko} \&  {Im}}{2005}]{ko05}
{Ko} J., {Im} M., 2005, J. Korean Astron. Soc., 38, 149

\bibitem[\protect\citeauthoryear{{La Barbera} et~al.}{{La Barbera}
    et~al.}{2010}]{lb10}
{La Barbera} F., {de Carvalho} R.~R., {de la Rosa} I.~G., {Gal} R.~R.,
{Swindle} R., {Lopes} P.~A.~A., 2010, arXiv1006.4056

\bibitem[\protect\citeauthoryear{{La Barbera} \& {de Carvalho}}{{La
      Barbera} \&  {de Carvalho}}{2009}]{lb09}
{La Barbera} F.,  {de Carvalho} R. R.,  2009, \apjl, 699, L76

\bibitem[\protect\citeauthoryear{{La Barbera} et~al.}{{La Barbera}
    et~al.}{2005}]{lb05}
{La Barbera} F., {de Carvalho} R.~R., {Gal} R.~R., {Busarello} G.,
{Merluzzi} P., {Capaccioli} M.,	{Djorgovski} S.~G., 2005, \apjl, 626,
19

\bibitem[\protect\citeauthoryear{{La Barbera} et~al.}{{La Barbera}
    et~al.}{2003}]{lb03}
{La Barbera} F., {Busarello} G., {Massarotti} M., {Merluzzi} P.,
{Mercurio} A., 2003, \aap, 409, 21L


\bibitem[\protect\citeauthoryear{{Larson}}{{Larson}}{1974}]{larsonSN:74}
{Larson} R.~B.,  1974, MNRAS, 169, 229

\bibitem[\protect\citeauthoryear{{Lee}, {Lee}, {Park} \& {Choi}}{{Lee}
  et~al.}{2008}]{lee08}
{Lee} J.~H.,  {Lee} M.~G.,  {Park} C.,    {Choi} Y.-Y.,  2008, \mnras, 389,
  1791

\bibitem[\protect\citeauthoryear{{Li} \& {Han}}{{Li} \&
    {Han}}{2007}]{li07}
{Li} Z., {Han} Z., 2007, \aap, 471, 795

\bibitem[\protect\citeauthoryear{{MacArthur}, {Gonz{\'a}lez} \&
  {Courteau}}{{MacArthur} et~al.}{2009}]{macarthur09}
{MacArthur} L.~A.,  {Gonz{\'a}lez} J.~J.,    {Courteau} S.,  2009, \mnras, 395,
  28

\bibitem[\protect\citeauthoryear{{Menanteau}, {Abraham} \& {Ellis}}{{Menanteau}
  et~al.}{2001}]{menanteau:01}
{Menanteau} F.,  {Abraham} R.~G.,    {Ellis} R.~S.,  2001, MNRAS, 322, 1

\bibitem[\protect\citeauthoryear{{Menanteau}}{{Menanteau} {et~al.}}{2005}]{menanteau:05}
{Menanteau} F.,  {et~al.} 2005, APJ, 620, 697

\bibitem[\protect\citeauthoryear{{Michard}}{{Michard}}{2005}]{michard:05}
{Michard} R.,  2005, \aap, 441, 451

\bibitem[\protect\citeauthoryear{{Oke \& Sandage}}{{Oke \& Sandage}}{1968}]{oke68}
{Oke}, J.~B. and {Sandage}, A., 1968, \apj, 154, 210

\bibitem[\protect\citeauthoryear{{Park} \& {Choi}}{{Park} \&
  {Choi}}{2005}]{park05}
{Park} C.,  {Choi} Y.-Y.,  2005, \apjl, 635, L29

\bibitem[\protect\citeauthoryear{{Peletier}, {Davies}, {Illingworth}, {Davis}
  \& {Cawson}}{{Peletier} et~al.}{1990}]{peletier:90}
{Peletier} R.~F.,  {Davies} R.~L.,  {Illingworth} G.~D.,  {Davis} L.~E.,
  {Cawson} M.,  1990, AJ, 100, 1091

\bibitem[\protect\citeauthoryear{{Pipino}, {D'Ercole}, {Chiappini} \&
  {Matteucci}}{{Pipino} et~al.}{2010}]{pipino10}
{Pipino} A.,  {D'Ercole} A.,  {Chiappini} C.,    {Matteucci} F.,  2010, ArXiv
  e-prints

\bibitem[\protect\citeauthoryear{{Poggianti}}{{Poggianti}}{1997}]{poggianti97}
{Poggianti} B.~M.,  1997, \aaps, 122, 399

\bibitem[\protect\citeauthoryear{{Rana} \& {Basu}}{{Rana} \&
  {Basu}}{1992}]{rb_imf}
{Rana} N.~C.,  {Basu} S.,  1992, \aap, 265, 499

\bibitem[\protect\citeauthoryear{{Rawle}, {Smith} \& {Lucey}}{{Rawle}
  et~al.}{2010}]{tim10}
{Rawle} T.~D.,  {Smith} R.~J.,    {Lucey} J.~R.,  2010, \mnras, 401, 852

\bibitem[\protect\citeauthoryear{{Roche}, {Bernardi} \& {Hyde}}{{Roche}
  et~al.}{2009}]{roche10}
{Roche} N.,  {Bernardi} M.,    {Hyde} J.,  2009, \mnras, tmp 947

\bibitem[\protect\citeauthoryear{{Roche}, {Bernardi} \& {Hyde}}{{Roche}
  et~al.}{2009}]{roche09}
{Roche} N.,  {Bernardi} M.,    {Hyde} J.,  2009, \mnras, 398, 1549

\bibitem[\protect\citeauthoryear{{S{\'a}nchez-Bl{\'a}zquez}, {Courty}, {Gibson}
  \& {Brook}}{{S{\'a}nchez-Bl{\'a}zquez} et~al.}{2009}]{sb09}
{S{\'a}nchez-Bl{\'a}zquez} P.,  {Courty} S.,  {Gibson} B.~K.,    {Brook} C.~B.,
   2009, \mnras, 398, 591

\bibitem[\protect\citeauthoryear{{S{\'a}nchez-Bl{\'a}zquez}, {Gorgas} \&
  {Cardiel}}{{S{\'a}nchez-Bl{\'a}zquez} et~al.}{2006}]{sanchez-blazquez}
{S{\'a}nchez-Bl{\'a}zquez} P.,  {Gorgas} J.,    {Cardiel} N.,  2006, \aap, 457,
  823

\bibitem[\protect\citeauthoryear{{Schawinski} {et~al.}}{{Schawinski}
 {et~al.}}{2009}]{zoo:blue}
{Schawinski} K.,  {et~al.} 2009, \mnras, 396, 818

\bibitem[\protect\citeauthoryear{{Scodeggio}}{{Scodeggio}}{2001}]{scodeggio01}
{Scodeggio}, M. 2001, \aj, 121, 2413

\bibitem[\protect\citeauthoryear{{Shimasaku}}{{Shimasaku}}{2001}]{shimasaku:01}
{Shimasaku} K.,  2001, AJ, 122, 1238


\bibitem[\protect\citeauthoryear{{Spergel} {et.~al.}}{{Spergel} {et.~al.}}{2003}]{wmap1}
{Spergel} D.~N. {et.~al.}, 2003, \apjs, 148, 175S

\bibitem[\protect\citeauthoryear{{Spolaor}, {Proctor}, {Forbes} \&
  {Couch}}{{Spolaor} et~al.}{2009}]{spolaor09}
{Spolaor} M.,  {Proctor} R.~N.,  {Forbes} D.~A.,    {Couch} W.~J.,  2009,
  \apjl, 691, L138

\bibitem[\protect\citeauthoryear{{Springel} \& {Hernquist}}{{Springel} \&
  {Hernquist}}{2003}]{springel03}
{Springel} V.,  {Hernquist} L.,  2003, \mnras, 339, 289

\bibitem[\protect\citeauthoryear{{Strateva}}{{Strateva}}{2001}]{strateva:01}
{Strateva} I.,  2001, AJ, 122, 1861

\bibitem[\protect\citeauthoryear{{Strauss et~al.}}{{Strauss et~al.}}{2002}]{strauss:02}
{Strauss} M.~A.,  {et~al.} 2002, AJ, 124, 1810

\bibitem[\protect\citeauthoryear{{Stringer} \& {Benson}}{{Stringer} \&
  {Benson}}{2007}]{stringer07}
{Stringer} M.~J.,  {Benson} A.~J.,  2007, \mnras, 382, 641

\bibitem[\protect\citeauthoryear{{Suh}, {Jeong}, {Oh}, {Yi}, {Ferreras} \&
  {Schawinski}}{{Suh} et~al.}{2010}]{suh10}
{Suh} H.,  {Jeong} H.,  {Oh} K.,  {Yi} S.~K.,  {Ferreras} I.,    {Schawinski}
  K.,  2010, \apjs, 187, 374

\bibitem[\protect\citeauthoryear{{Tamura} \& {Ohta}}{{Tamura} \&
  {Ohta}}{2003}]{tamura03}
{Tamura} N.,  {Ohta} K.,  2003, AJ, 126, 596

\bibitem[\protect\citeauthoryear{{Tamura} \& {Ohta}}{{Tamura} \&
  {Ohta}}{2004}]{tamura04}
{Tamura} N.,  {Ohta} K.,  2004, MNRAS, 355, 617

\bibitem[\protect\citeauthoryear{{Tortora}, {Napolitano}, {Cardone},
  {Capaccioli}, {Jetzer} \& {Molinaro}}{{Tortora} et~al.}{2010}]{tortora10}
{Tortora} C.,  {Napolitano} N.~R.,  {Cardone} V.~F.,  {Capaccioli} M.,
  {Jetzer} P.,    {Molinaro} R.,  2010, ArXiv e-prints

\bibitem[\protect\citeauthoryear{{Vlaji{\'c}}, {Bland-Hawthorn} \&
  {Freeman}}{{Vlaji{\'c}} et~al.}{2009}]{vlajic09}
{Vlaji{\'c}} M.,  {Bland-Hawthorn} J.,    {Freeman} K.~C.,  2009, \apj, 697,
  361

\bibitem[\protect\citeauthoryear{{Weinmann}, {Kauffmann}, {van den Bosch},
  {Pasquali}, {McIntosh}, {Mo}, {Yang} \& {Guo}}{{Weinmann}
  et~al.}{2009}]{weinmann09}
{Weinmann} S.~M.,  {Kauffmann} G.,  {van den Bosch} F.~C.,  {Pasquali} A.,
  {McIntosh} D.~H.,  {Mo} H.,  {Yang} X.,    {Guo} Y.,  2009, \mnras, 394, 1213

\bibitem[\protect\citeauthoryear{{Worthey}}{{Worthey}}{1994}]{worthey94}
{Worthey} G.,  1994, \apjs, 95, 107

\bibitem[\protect\citeauthoryear{{Wu}, {Shao}, {Mo}, {Xia} \& {Deng}}{{Wu}
  et~al.}{2005}]{wu:05}
{Wu} H.,  {Shao} Z.,  {Mo} H.~J.,  {Xia} X.,    {Deng} Z.,  2005, \apj, 622,
  244

\bibitem[\protect\citeauthoryear{{Yamauchi} \& {Goto}}{{Yamauchi} \&
  {Goto}}{2005}]{yamauchi05}
{Yamauchi} C.,  {Goto} T.,  2005, \mnras, 359, 1557

\bibitem[\protect\citeauthoryear{{York}, {Anderson} Jr., {Anderson} \& {SDSS
  collaboration}}{{York} {et~al.}}{2000}]{york:00}
{York} D.~G.,  {Anderson} Jr. J.~E.,  {Anderson} S.~F.,    {SDSS collaboration}
  2000, AJ, 120, 1579

\end{thebibliography}

\appendix
\section{Possible systematic effects in the colour gradients}\label{ap-sys}

\subsection{Effect of the ellipticity}

\begin{figure}
\includegraphics[width=9.cm]{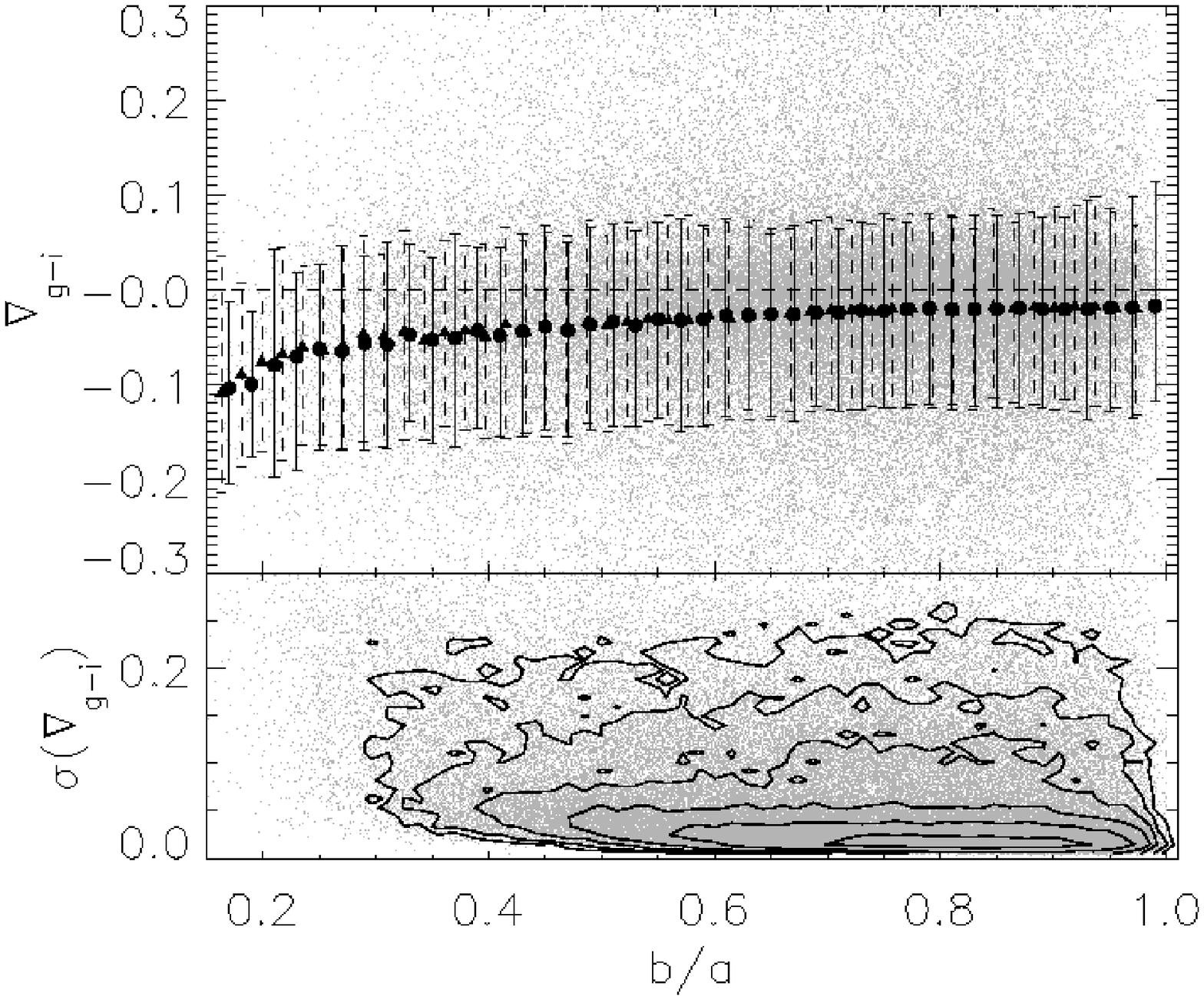}
\caption{ Variation of $(g-i)$ colour gradients with
  $b/a$ in the r-band, top panel, and its errors, bottom panel, for
    galaxies within the {\it S20.5} sample. In the top panel we have
    superimposed median colour gradients values and their $1\sigma$
    dispersion range for $b/a$ in the r-band, circles with solid error bars, but also for
    $b/a$ in the g-band, triangles with dashed error bars. The bottom panel shows density contours with increments of a factor of two.}
\label{fig:elip_med}
\end{figure}

We obtain colour gradients from surface brightness measurements
  that are circularly averaged. Thus, colour gradients within
  elongated galaxies could be diluted. To address this issue we have
  separated galaxies within sample {\it S20.5}, the most populated
  one, into different ranges according to the ratio between the
  projected minor and major axes of the galaxies in the r-band,
  $b/a$. The top panel of Fig. \ref{fig:elip_med} shows the variation
  of the $(g-i)$ colour gradient with $b/a$. We can see that more
  elongated galaxies present steeper colour gradients, with median
  values varying from $-0.07$ for galaxies with $b/a< 0.4$ to $-0.04$
  for $b/a> 0.8$. This tendency appears independently of the band used
  to define $b/a$. The same tendency of steeper gradient for more elongated galaxies is found for galaxies in the
  other studied samples. It is also seen when colour gradients are
  defined as the variation of colour per logarithmic normalised
  radius. However, this is not observed for the  $(r-z)$
  colour gradients, where all median values are similar, $\sim-0.07$,
  independently of their $b/a$ value . 

The bottom panel of Fig. \ref{fig:elip_med} shows that the distribution of the error does not appreciably change as a function the projected
axis ratio. Thus, the error distribution is not enough to explain the slope change of the median values around
$b/a< 0.3$ seen in the top panel of Fig. \ref{fig:elip_med}.

This result shows that the dilution of colour gradients due to the
use of circularly averaged surface brightness does not erase colour
gradients although it may still dilute them somewhat. 

Galaxies in the range $b/a< 0.4$ are late-types. Very inclined
late-type galaxies are most affected by their internal absorption,
something that could explain their steeper colour gradients. Due to
their small gas content, early-type galaxies properties are not expected to be affected by their inclination. 

\citet{choi07} studied the dimming effect for inclined late-type
galaxies finding that it is better to exclude those with $b/a< 0.6$
when studying properties that could be affected by galactic internal
dimming. Considering only galaxies with $b/a> 0.6$, we find that the
linear variation of the median of the $(g-i)$ colour gradient with
$b/a$ has a slope smaller than $0.02$, for all of the
luminosity-threshold samples considered in our study. As expected from
the large dispersion found in Fig. \ref{fig:elip_med}, the Spearman
coefficient is too small to support statistically such a linear
relation. 

We can conclude that colour gradients do not vary
in a significant manner with $b/a$, except for very inclined
late-type galaxies, which are most likely affected by their internal
dimming. Since the z-band is the least affected by dust extinction,
this last result is consistent with the lack of $(r-z)$ colour
gradient variation with $b/a$.

\subsection{Effect of the number of annuli considered}\label{a_rad}

\begin{figure}
{\epsfxsize=8.5truecm \epsfbox[57 8 561 473]{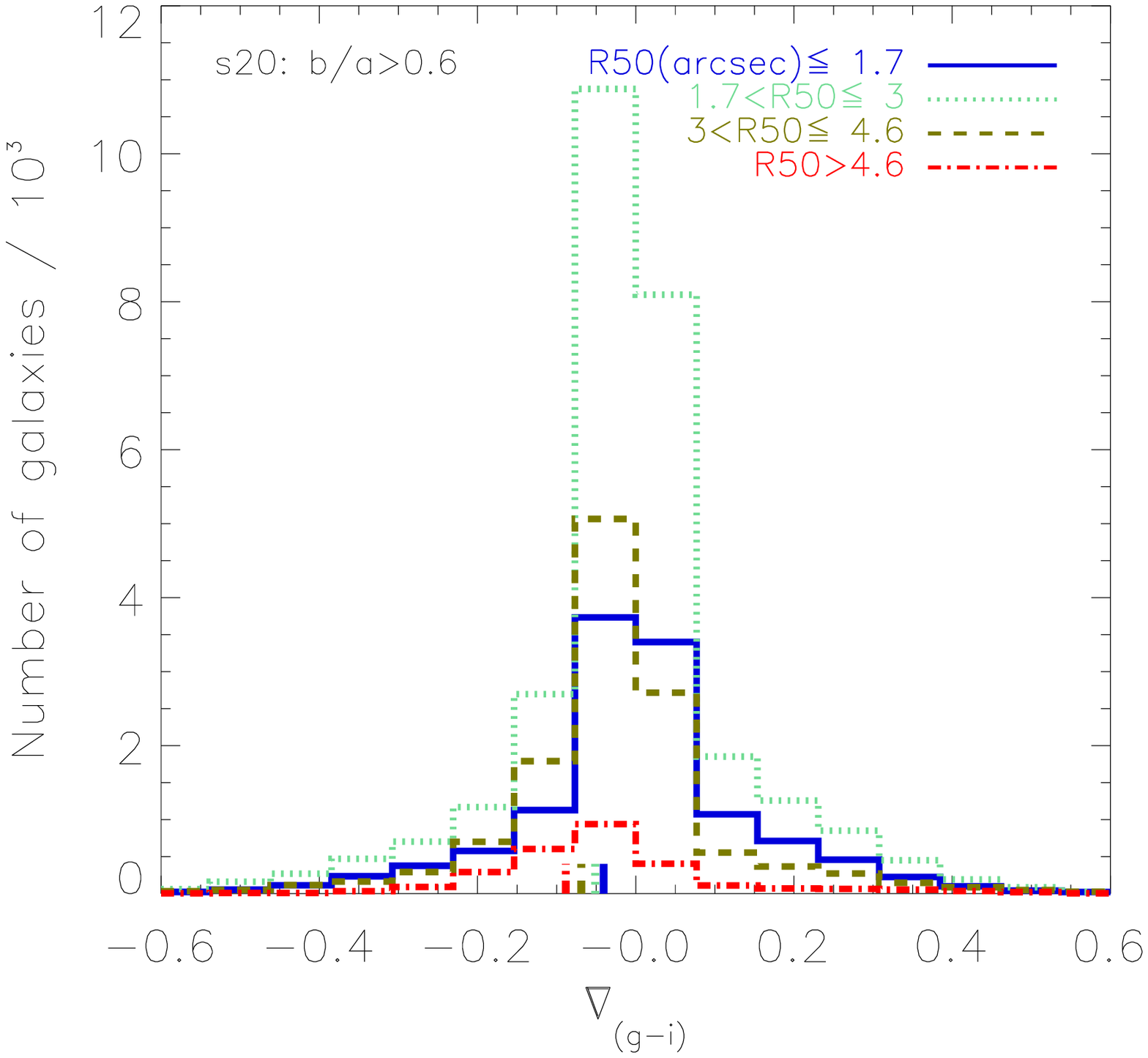}}
\caption{ Distribution of $(g-i)$ colour gradients for galaxies with
  $b/a > 0.6$, within the {\it S20} sample separated by the radius
  containing half their light: $R50 \le 1.7$ arcsec (outer radius of
  $4^{th}$ annulus, solid line), $1.7 <R50\, (arcsec) \le 3$ (limits of
  the $5^{th}$ annulus, dotted line), $3<R50\, (arcsec)\le 4.6$
  (limits of the $6^{th}$ annulus, dashed line), $R50>4.6$ arcsec (outside the $6^{th}$ annulus, dot-dashed line). }
\label{fig:r50_arcsec}
\end{figure}

Here we study the bias that we could be introducing by leaving out the
innermost light, $R<0.67$ arcsec, of the galaxy. For all the studied
  galaxies, colour gradients have
  been obtained from the surface brightness within two times the
  radius containing half the total galactic light. In section
\S\ref{sec:dr7sample} it was pointed out that the {\it S21}
sample contains galaxies with half light radii, $R50$, within the two
innermost annuli that we neglect computing the colour gradient. For
the rest of the samples this does not occur. Moreover, in these
samples less than $5$\% of the galaxies have their half light radii
within the first annuli considered in the calculation of colour
gradients, $0.67<R50\, (arcsec)<1.0$  ($3^{rd}$ annulus in the SDSS database).

Fig. \ref{fig:r50_arcsec} shows the distribution of colour gradients
of galaxies with $b/a>0.6$ from the {\it S20} sample, divided in
four ranges of half light radii. The cut in ellipticity
allows us to isolate the tendency of colour gradients with $R50$. The
ranges of $R50$ have been defined using as boundaries the outer radii of the $4^{th}$, $5^{th}$ and $6^{th}$ annuli in which the SDSS database provides averaged surface brightness measurements. Fig. \ref{fig:r50_arcsec} shows that the largest amount of galaxies in sample {\it S20} have their $R50$ within the $5^{th}$ surface brightness annuli. This is also the case when considering all galaxies, independently of their ellipticity and it also holds for galaxies within the sample {\it S20.5}, while for sample {\it S19} there is an even amount of galaxies with their $R50$ in either the $5^{th}$ or $6^{th}$ annulus. As it was pointed out before, all galaxies in our samples have at least four surface brightness measurements, from the $3^{rd}$ to the $6^{th}$ annulus.  Thus, except for the {\it S21} sample, when obtaining colour gradients we are taking into account surface brightness measurements both inside and outside $R50$. This point is essential for any robust conclusion about the behaviour of colour gradients.

\begin{figure}
\includegraphics[width=9.cm]{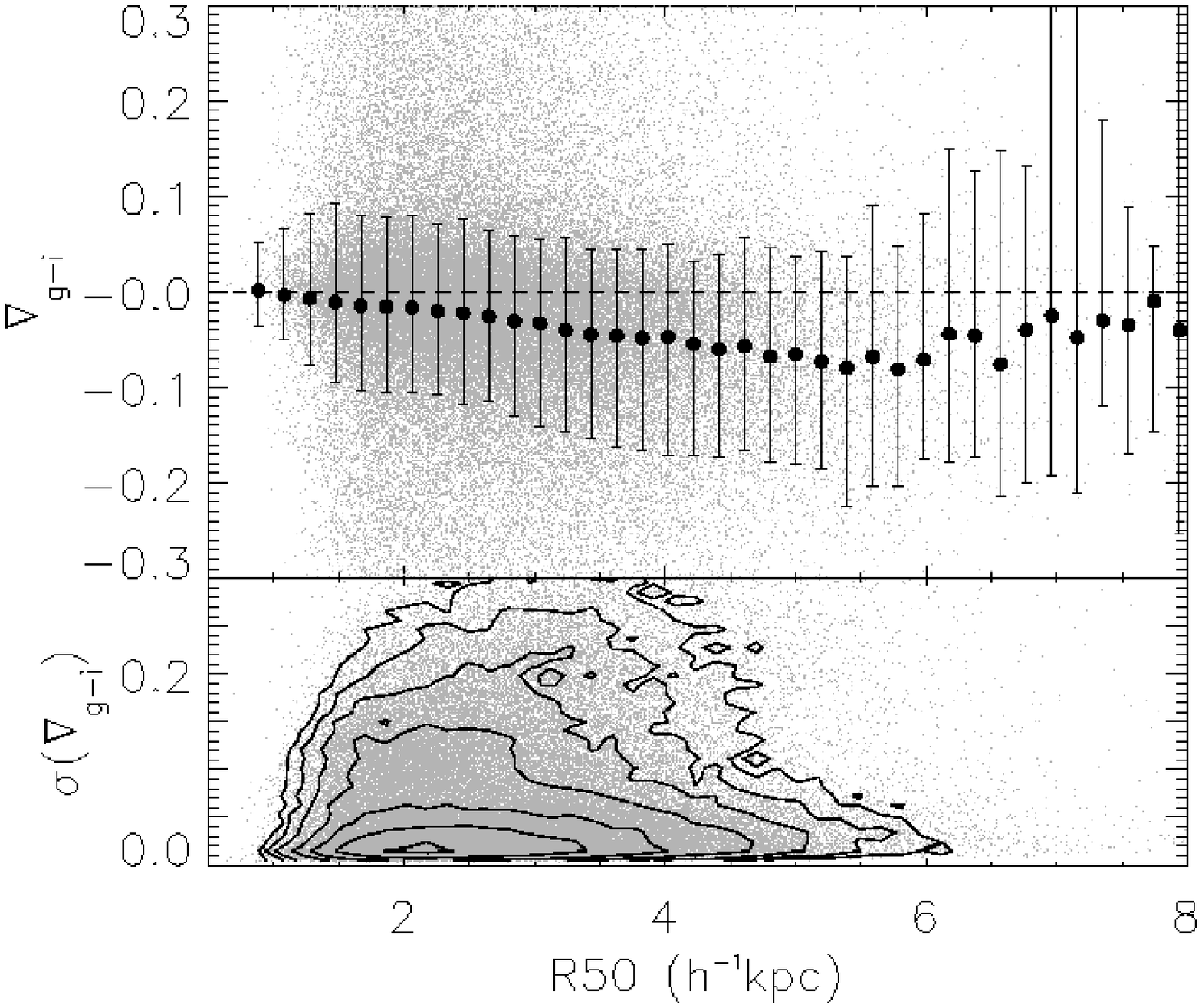}
\caption{ Similar to Fig. \ref{fig:elip_med} for the radius
    containing half the total light ($h^{-1}kpc$)  within galaxies from the {\it S20} sample.}
\label{fig:r50_kpc}
\end{figure}

Median colour gradients are steeper for higher values of $R50\,
(arcsec)$, for all the defined luminosity-threshold samples. The same tendency has been found in the
  study by \citet{roche10} on early-type galaxies.

In Fig. \ref{fig:r50_arcsec} the ranges of values we study are defined
by half light radii values in $arcsec$. Therefore, gradient values could be
smeared out depending on the amount of light considered. To further
explore this point, we have also studied the tendency of colour
gradients when comparing galaxies with similar physical transverse
radius, i.e., defining ranges of $R50$ measured in
$h^{-1}kpc$. Different amount of surface brightness annuli account for
a different amount of light for galaxies with similar sizes at
different distances. Thus, in this way galaxies with colour gradients
extracted from different amount of annuli are included in the same
range.

Independently of the galactic ellipticity, the same tendency found for
$R50 (arcsec)$ applies when separating galaxies by their $R50
(h^{-1}kpc)$: the larger $R50$ is, the steeper a colour gradient is
likely to be. Galaxies within the {\it S20.5} sample have median
$(g-i)$ colour gradients: $-0.036$, $-0.048$, $-0.061$, $ -0.080$ for
$R50 (h^{-1}kpc)\leq 2$; $2<R50(h^{-1}kpc)\leq 3$;
$3<R50(h^{-1}kpc)\leq 4$; $R50 (h^{-1}kpc)> 4$, respectively. The same
tendency for median values is found for galaxies within the {\it S19}
and {\it S20} samples. In Fig. \ref{fig:r50_kpc} we present the
variation with $R50 (h^{-1}kpc)$ of the $(g-i)$ colour gradient and
its error, for galaxies in {\it S20}. Median colour gradient
  values tend to decrease with larger half light radii, though, the
  correlation is dominated by the dispersion. When defining the colour
  gradient in decades of normalised radii (radius in logarithmic
  units), we also find that colour gradients get steeper for galaxies
  with larger $R50$. \citet{tortora10} finds a similar tendency with
  effective radius, supporting the robustness of our result.

Dividing our sample into late and early-types according to the
concentration index, we found that the colour-gradient of late-type
galaxies change more rapidly with the galactic size than
early-types. The corresponding slopes for samples {\it S19}, {\it S20}
and {\it S25} are: $-0.028$, $-0.025$ and $-0.024$ for late-types and
$-0.004$, $-0.002$ and $-0.002$ for early-types galaxies. The colour
gradient of early-type galaxies, remains quite constant with galactic size.

For the samples of galaxies under study, the colour gradient variation
with $R50$ has been found to be independent of the amount of light
used for obtaining it. Still, the deeming of colour gradients could be
partly due to calculating them with fewer annuli with surface
brightness measurements. If this was the case, we would expect a
correlation between the gradient error and
$R50(arcsec)$. However, no clear tendency is found, though the
errors dispersion clearly decreases for larger $R50$ (see Fig. \ref{fig:r50_arcsec}). This tendency
cannot explain alone the variation with $R50$ seen for colour
gradients. However, this consideration brings us to point out the
unclear role that colour gradients could play in biasing global
galactic properties obtained by extrapolating spectral characteristics
from the SDSS galactic cores. In their studies, both
\citet{scodeggio01} and \citet{roche09} found out that the colour
gradients should be taken into account for estimating the evolution of
the colour-magnitude relation in early-type galaxies.

Despite the large dispersion in colour gradients values, the above results
suggest that on average, galaxies with larger $R50$, measured in
either $arcsec$ or $kpc$, intrinsically tend to have steeper colour
gradients.

\subsection{Effect of having close pairs of galaxies}

\begin{figure}
\includegraphics[width=9.1cm]{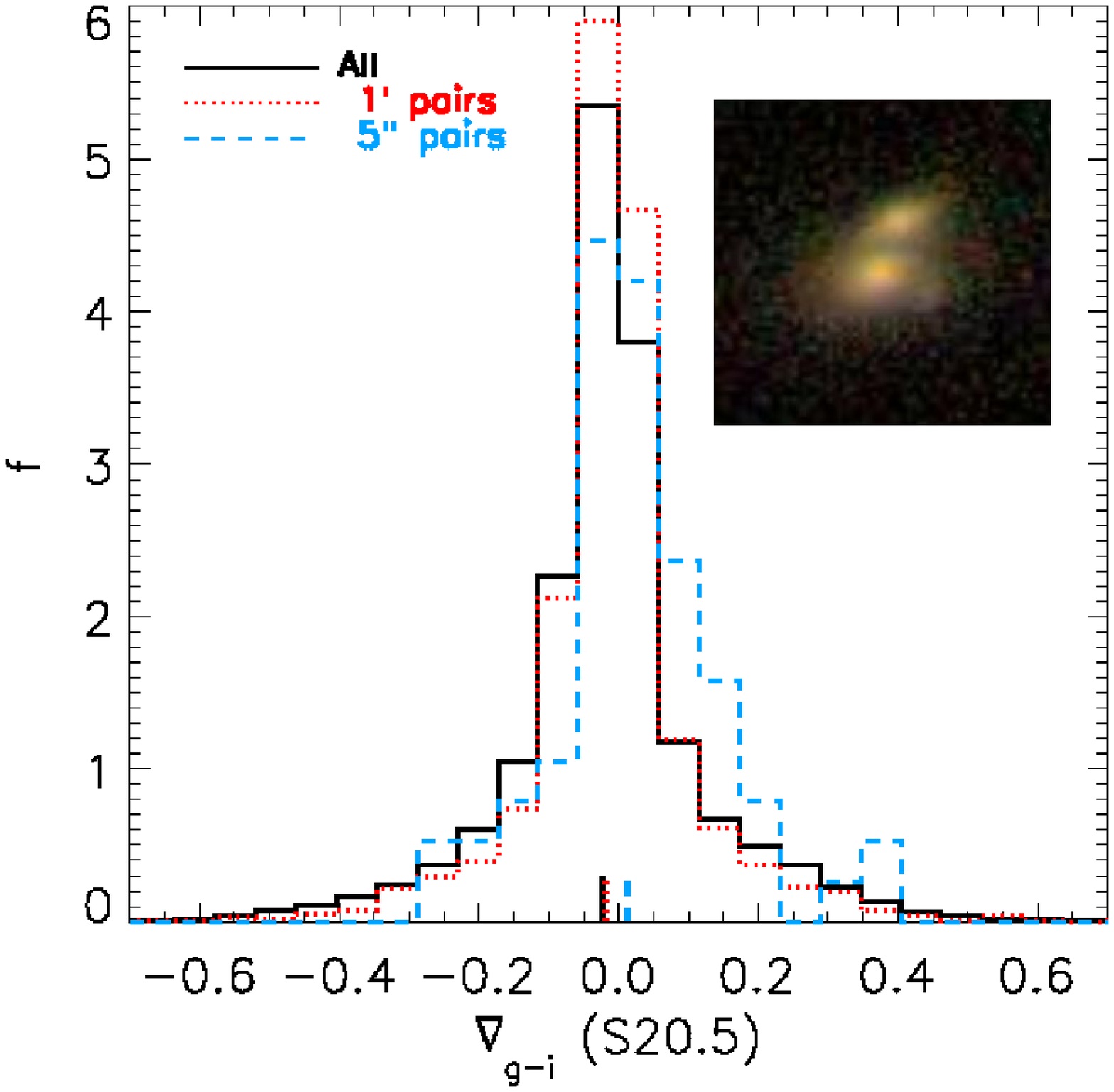}
\caption{ Distribution of (g-i) colour gradient for all the
  galaxies within the {\it S20.5} sample, solid line, and for only close
  pairs of galaxies within either $1$ arcmin, dotted line, or $5$ arcsec, dashed line, projected radius and
  $100\, km/s$ in redshift. The inset shows an
    example of a pair of galaxies pair closer than $5$ arcsec projected
    radius. Histograms are normalised to give unit area underneath.}
\label{fig:pairs}
\end{figure}


Interacting galaxies can affect the colour gradients of each
other \citep[see e.g.][]{ko05}. Moreover, colour gradient can be ill defined for close pairs of
galaxies. Besides, the sky subtraction for close pairs of galaxies
within the SDSS database may be problematic (John Lucey, private
communication). Here we explore how the colour gradient distribute for
close pairs of galaxies within our defined luminosity-threshold
samples.

Fig. \ref{fig:pairs} show the distribution of colour gradients for
galaxies that are within $100\, km/s$ in redshift and within a
projected radius of either $1$ arcmin or $5$ arcsec. This constitutes a lower
limit to the fraction of close galactic pairs, since this number is
expected to increase when including the whole photometric SDSS
sample.\footnote{SDSS could not observe galaxies closer than 55 arcsec in
  the same plate and therefore close pairs are under represented in
  the spectroscopic sample. They were only observed in regions where
  spectroscopic plates overlapped.}

We can observe that those close pairs with projected separations of
maximum $1$ arcmin, present a distribution with a median value coincident
with that of the total sample. These are only $5$\% of the total {\it
  S20.5} sample. Within the studied redshift range, $1$' corresponds
to around $100\, kpc$, approximately half the typical Virial radius of
galaxies as massive as the Milky Way \citep{acooper09}. Thus, this
will be an approximate limit to find interacting pairs of galaxies,
which could affect each other colour gradients. Since no clear bias is
found we do not have any further reason to consider that the colour
gradients of close pairs of galaxies as far as $1$ arcmin apart is ill defined.

In the {\it S20.5} sample, we find $128$ galaxies that are within  $100\, km/s$ in redshift and
$5$ arcsec projected from another galaxy. A visual inspection of these
galaxies reveals pairs of interacting galaxies, with some of them
showing clear tidal tails. However, most of these close pairs appear just
as red double nucleated structures, similar to that shown in the inset
of Fig. \ref{fig:pairs}. 

It is very interesting to see in
Fig. \ref{fig:pairs} that these galaxies tend to present bluer cores
than what expected from the total sample. This result could point out
that the star formation triggered by galactic interactions could be
responsible for the observed bluer cores. Simulation such as the one from \citet{kobayashi:04} suggest that this phase will decay with time into a flatter colour gradient. In any case, these galaxies represent a percentage
far below $1$\% of the initial {\it
  S20.5} sample, and, therefore their contribution to increase the
presence of blue nucleated galaxies is negligible at least in our sample.

\end{document}